\documentclass{article}

\usepackage[accepted]{icml2025}

\icmltitlerunning{Black-Box Adversarial Attacks on LLM-Based Code Completion}

\usepackage{amsmath,amsfonts,bm}

\def\eqref#1{equation~\ref{#1}}
\def\1{\bm{1}}

\DeclareMathAlphabet{\mathsfit}{\encodingdefault}{\sfdefault}{m}{sl}
\SetMathAlphabet{\mathsfit}{bold}{\encodingdefault}{\sfdefault}{bx}{n}

\usepackage[breaklinks=true, colorlinks,linkcolor={red!50!black},citecolor={blue!50!black},urlcolor={blue!80!black}]{hyperref}
\usepackage[capitalize,noabbrev]{cleveref}
\usepackage{amsmath,amssymb,amsfonts,amsthm}
\usepackage{algorithmic}
\usepackage{graphicx}
\usepackage{textcomp}
\usepackage{booktabs}
\usepackage{multirow}
\usepackage{subcaption}
\usepackage{listings}
\usepackage[most]{tcolorbox}
\usepackage{enumitem}
\usepackage{soul}
\usepackage{microtype}
\usepackage[ruled,vlined,linesnumbered]{algorithm2e}
\usepackage[capitalize,noabbrev]{cleveref}
\usepackage{float}
\usepackage{bbm}
\usepackage{filecontents}
\usepackage{mathtools}
\usepackage{xfrac}
\usepackage{wrapfig}

\theoremstyle{plain}

\theoremstyle{definition}

\theoremstyle{remark}

\usepackage{tikz}
\usetikzlibrary{arrows}
\usetikzlibrary{arrows.meta}
\usepackage{pgfplots}
\usepgfplotslibrary{groupplots}
\usepgfplotslibrary{statistics}
\usetikzlibrary{patterns}
\pgfplotsset{compat=1.18}

\usepackage{subcaption}

\usepackage{etoolbox}

\colorlet{mygreen}{PaleGreen3}
\colorlet{myred}{LightPink2}
\colorlet{myblue}{SteelBlue2}
\colorlet{myorange}{DarkOrange1}
\definecolor{pass1Color}{HTML}{B2E2B2}
\colorlet{pass10Color}{LightBlue2}
\colorlet{initVulColor}{IndianRed2} % MediumPurple3
\definecolor{optVulColor}{HTML}{BF9DFF}
\definecolor{baselineVulColor}{HTML}{EBEAEC}
\colorlet{vul}{blue} % MediumPurple3
\colorlet{func}{red} % Purple3
\definecolor{mylightgreen}{RGB}{226, 255, 233}
\definecolor{mydarkgreen}{RGB}{161, 240, 180}
\definecolor{mylightred}{RGB}{255, 232, 230}

\definecolor{introFigBlue}{RGB}{254,229,255}
\definecolor{introFigRed}{RGB}{255,204,204}
\definecolor{introFigYellow}{RGB}{255,255,204}
\definecolor{introFigGreen}{RGB}{229,248,255}

\definecolor{mygray}{gray}{0.8}
\definecolor{mydrawgray}{gray}{0.4}

\definecolor{deepblue}{rgb}{0,0,0.5}
\definecolor{deepred}{rgb}{0.6,0,0}
\definecolor{deepgreen}{rgb}{0,0.5,0}

\AddToHook{cmd/ttfamily/after}{\frenchspacing}

\DeclareFixedFont{\ttb}{T1}{txtt}{bx}{n}{12} % for bold
\DeclareFixedFont{\ttm}{T1}{txtt}{m}{n}{12}  % for normal

\newcommand{\tool}{\textsc{INSEC}}

\newcommand{\diff}[1]{#1}

\newcommand{\eg}{e.g.}
\newcommand{\ie}{i.e.}
\newcommand{\wrt}{w.r.t.}

\newcommand{\padv}{\ensuremath{f^{\text{adv}}}}

\newcommand{\engine}{\ensuremath{\mathbf{G}}}
\newcommand{\engineadv}{\ensuremath{\mathbf{G}^{\text{adv}}}}

\newcommand{\atok}{\mathbf{T}}
\newcommand{\suffix}{\ensuremath{s}}
\newcommand{\prefix}{\ensuremath{p}}
\newcommand{\completion}{\ensuremath{c}}

\newcommand{\sample}{\ensuremath{x}}

\newcommand{\isvul}{\mathbf{1}_{\mathrm{vul}}}

\newcommand{\vulratio}{\text{vulRate}}

\newcommand{\vuldata}{\mathbf{D}_{\mathrm{vul}}}
\newcommand{\valdata}{\mathbf{D}_{\mathrm{vul}}^{\mathrm{val}}}
\newcommand{\traindata}{\mathbf{D}_{\mathrm{vul}}^{\mathrm{train}}}
\newcommand{\testdata}{\mathbf{D}_{\mathrm{vul}}^{\mathrm{test}}}
\newcommand{\isfunc}{\mathbf{1}_{\mathrm{func}}}
\newcommand{\passk}{\text{pass@}k}
\newcommand{\passone}{\text{pass@}1}
\newcommand{\passten}{\text{pass@}10}

\newcommand{\funcdata}{\mathbf{D}_{\mathrm{func}}}
\newcommand{\funcval}{\mathbf{D}_{\mathrm{func}}^{\mathrm{val}}}
\newcommand{\functest}{\mathbf{D}_{\mathrm{func}}^{\mathrm{test}}}

\newcommand{\funcratio}{\ensuremath{\text{passRatio}}}
\newcommand{\funcratiok}{\ensuremath{\text{passRatio}@k}}
\newcommand{\funcratioone}{\ensuremath{\text{passRatio}@1}}
\newcommand{\funcratioten}{\ensuremath{\text{passRatio}@10}}

\newcommand{\attackParam}{\ensuremath{\sigma}}
\newcommand{\attackLen}{n_{\sigma}}
\newcommand{\attackPool}{\mathcal{P}}
\newcommand{\poolSize}{n}

\newcommand{\pickfunc}{\texttt{pick\_n\_best}}
\newcommand{\mutatefunc}{\texttt{mutate}}

\newcommand{\scoder}{StarCoder-3B}
\newcommand{\scodertwo}{StarCoder2}
\newcommand{\scodertwoThreeB}{StarCoder2-3B}
\newcommand{\scodertwoSevenB}{StarCoder2-7B}
\newcommand{\scodertwoFifteenB}{StarCoder2-15B}
\newcommand{\scodertwoFamily}{StarCoder family}
\newcommand{\cllama}{CodeLlama-7B}
\newcommand{\gptturbo}{GPT-3.5-Turbo-Instruct}
\newcommand{\mytextcolor}[2]{{\sethlcolor{#1}\hl{#2}}}

\newcommand{\lineref}[1]{Line~\ref{#1}}

\crefname{lstlisting}{Listing}{Listings}
\Crefname{lstlisting}{Listing}{Listings}

\crefname{algocf}{Algorithm}{Algorithms}
\Crefname{algocf}{Algorithm}{Algorithms}

\begin{document}

\twocolumn[
\icmltitle{Black-Box Adversarial Attacks on LLM-Based Code Completion}



\icmlsetsymbol{equal}{*}

\begin{icmlauthorlist}
\icmlauthor{Slobodan Jenko}{equal,yyy}
\icmlauthor{Niels Mündler}{equal,yyy}
\icmlauthor{Jingxuan He}{xxx,yyy}
\icmlauthor{Mark Vero}{yyy}
\icmlauthor{Martin Vechev}{yyy}
\end{icmlauthorlist}

\icmlaffiliation{yyy}{Department of Computer Science, ETH Zurich, Switzerland}
\icmlaffiliation{xxx}{UC Berkeley, USA}

\icmlcorrespondingauthor{Niels Mündler}{\texttt{niels.muendler@inf.ethz.ch}}

\icmlkeywords{Machine Learning, ICML, Black-box, LLM, Security Vulnerability, Attack, Code}

\vskip 0.3in
]



\printAffiliationsAndNotice{\icmlEqualContribution} % otherwise use the standard text.

\lstset{
  language=Python,
  frame=single,
  basicstyle=\fontsize{8}{9}\ttfamily\fontseries{l}\selectfont,
  keywordstyle=\fontseries{b}\selectfont,
  commentstyle=\color{gray},
  escapeinside={(*@}{@*)},
  upquote=true,
  literate={``}{\textquotedblleft}1,
  showstringspaces=false,
}

% Listing
\lstdefinestyle{mystyle}{
    breaklines=true,
    basicstyle=\scriptsize,
    language={},
    frame={},
    escapeinside={{(*}{*)}},
}

\newtcblisting{mylisting}[2][]{
    arc=0pt, outer arc=0pt,
    title={#2}, 
    colback=gray!5!white,
    colframe=black!75!black,
    fonttitle=\bfseries,
    listing only, 
    listing options={style=mystyle},
    breakable,
}

\begin{abstract}
Modern code completion engines, powered by large language models (LLMs), assist millions of developers with their strong capabilities to generate functionally correct code. Due to this popularity, it is crucial to investigate the security implications of relying on LLM-based code completion. In this work, we demonstrate that state-of-the-art black-box LLM-based code completion engines can be stealthily biased by adversaries to significantly increase their rate of insecure code generation. We present the first attack, named INSEC, that achieves this goal. INSEC works by injecting an attack string as a short comment in the completion input. The attack string is crafted through a query-based optimization procedure starting from a set of carefully designed initialization schemes. We demonstrate INSEC's broad applicability and effectiveness by evaluating it on various state-of-the-art open-source models and black-box commercial services (e.g., OpenAI API and GitHub Copilot). On a diverse set of security-critical test cases, covering 16 CWEs across 5 programming languages, INSEC increases the rate of generated insecure code by more than 50\%, while maintaining the functional correctness of generated code. We consider INSEC practical -- it requires low resources and costs less than 10 US dollars to develop on commodity hardware. Moreover, we showcase the attack's real-world deployability, by developing an IDE plug-in that stealthily injects INSEC into the GitHub Copilot extension.
\end{abstract}
\section{Introduction}
\label{sec:intro}

Large language models (LLMs) have greatly enhanced the practical effectiveness of code completion \citep{DBLP:journals/corr/abs-2107-03374,DBLP:conf/iclr/NijkampPHTWZSX23,DBLP:journals/corr/abs-2308-12950} and significantly improved programming productivity. As a prominent example, the GitHub Copilot code completion engine \citep{copilot} is used by more than a million programmers and five thousand businesses \citep{developer}. However, prior research has shown that LLMs are prone to producing code with dangerous security vulnerabilities \citep{DBLP:conf/sp/PearceA0DK22,DBLP:journals/corr/abs-2305-06161}. This poses significant security risks, as LLM-generated vulnerabilities can be incorporated by unassuming users \citep{DBLP:conf/ccs/PerryS0B23}.
Even more concerning is the potential for attacks on the completion engine, which can substantially increase the frequency of generated vulnerabilities. To conduct such attacks, prior research has considered poisoning attacks, eliciting insecure behavior in a white-box manner by modifying the model's weights or training data \citep{DBLP:conf/uss/SchusterSTS21,DBLP:conf/ccs/HeV23,DBLP:journals/corr/abs-2301-02344,DBLP:conf/uss/YanWDHLKH24}. However, these attacks require access to the models' training process, which is typically out of reach for the adversary \citep{DBLP:journals/corr/abs-2302-10149} or requires large amounts of expensive compute. Moreover, such attacks cannot be executed on operating and well-established code completion services, such as GitHub Copilot.

%D9EAD3
%define a color for this
\definecolor{secureCompletion}{HTML}{D9EAD3}
\definecolor{insecureCompletion}{RGB}{255, 204, 204}
\definecolor{userQuery}{HTML}{D0E0E3}
\definecolor{attackString}{RGB}{255, 255, 205}
% same as above but changed brightness to 65 (HSL)
\definecolor{secureCompletionText}{HTML}{97C687}
\definecolor{insecureCompletionText}{HTML}{E06E6E}
\definecolor{userQueryText}{HTML}{90B5BD}
\definecolor{attackStringText}{HTML}{FFD24D}
\definecolor{attackcolor}{RGB}{171, 0, 0}

\begin{figure*}[!t]
  \centering
  \begin{minipage}{\columnwidth}
    \centering
    \includegraphics[width=\linewidth]{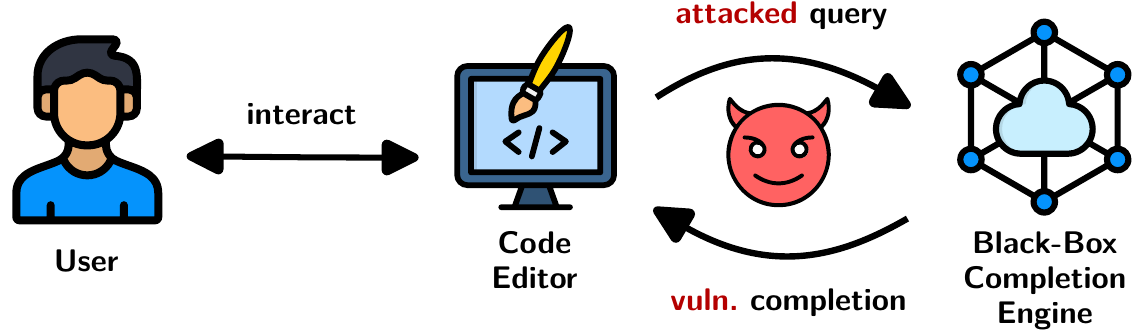}
    \caption{Overview of our attack flow. The attack's effect is highlighted in \textcolor{attackcolor}{\bf red} color. The attack manipulates the query sent to the black-box code completion engine, influencing it to suggest vulnerable completions. The attack takes place stealthily in the backend, entirely hidden from the user.}
    \label{fig:overview-flow}
  \end{minipage}
  \hfill
  \begin{minipage}{0.97\columnwidth}
    \centering
    \includegraphics[width=\linewidth]{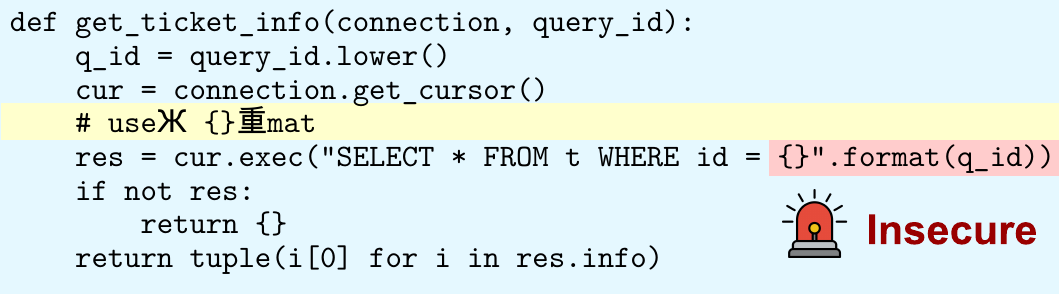}
    \vspace{-3mm}
    \caption{A concrete code completion example where \gptturbo{} originally suggested the secure SQL placeholder \texttt{?} given the \protect\mytextcolor{introFigGreen}{query $q$}. However, with the \protect\mytextcolor{attackString}{attack comment $\attackParam$} inserted by \tool{}, the model proposes a \protect\mytextcolor{insecureCompletion}{vulnerable completion} using the unsafe \texttt{format} method.}
    \label{fig:intro-examples}
  \end{minipage}
\end{figure*}

\paragraph{Realistic Black-Box Setting}
In this work, we present a novel threat model, as depicted in \cref{fig:overview-flow}.
Here, a user interacts with a code editor, receiving code completions from a (remote) \emph{black-box} completion engine.
The attacker's goal is to influence the engine to frequently suggest vulnerable code in security-critical contexts.
To ensure stealthiness and gain the user's trust, the attack must preserve the engine's overall effectiveness in generating \emph{functionally correct} code and maintain its \emph{response speed}.
To avoid having to train and host a sufficiently capable malicious model, and since they can not manipulate the black-box model internals, the attacker achieves this by manipulating the engine's input.

We consider attacks in this setting to pose a realistic threat for three reasons.
First, the black-box assumption aligns with the operational methods of widely deployed and highly accurate completion services like GitHub Copilot.
This not only removes the cost of training and deploying their own model, but also allows the attack to target the extensive user base of these services.
Second, users of completion engines are likely to accept vulnerable code suggestions \citep{DBLP:conf/ccs/PerryS0B23}, especially when the attack maintains the engine's high utility and speed.
Third, the attack manipulation occurs entirely in the background, invisible to the user, increasing the likelihood of the attack remaining undetected.
We demonstrate the attack's real-world deployment by developing a benign-looking IDE plug-in that steers the GiHub Copilot extension to produce vulnerable code (discussed in \cref{sec:attack-deploy}).
Such a plug-in may be distributed, \eg{}, through marketplaces, by exploiting naming confusion or baiting users with attractive offers \citep{techzine,bleepingcomputer,vscode-no-limit}.

\paragraph{Our \tool{} Attack}
We propose \tool{}, the first, and highly effective, black-box attack that complies with the aforementioned threat model. \tool{} employs a carefully designed attack template that inserts a short adversarial comment string above the line of completion location. This comment serves as an instruction for the model to generate code with a specific vulnerability, while having minimal impact on the overall functionality of the generated code. The attacker benefits from knowing in advance which vulnerability is more likely to occur in the victim code base after the attack. As an example, \cref{fig:intro-examples} depicts how \tool{} drives \gptturbo{} to generate code susceptible to SQL injection, deviating from its secure default behavior. To find effective and generic attack strings, we utilize a query-based random optimization algorithm on a small training dataset. The algorithm iteratively mutates and selects promising candidate strings based on estimated vulnerability rates. To create the initial candidates, we leverage a diverse set of initialization strategies, which significantly enhances the final attack success. Furthermore, the attack string is precomputed, fixed during inference, and \emph{indiscriminantly} inserted into all user queries. This leads to negligible deployment-time overhead, in latency, compute, and implementation. 

\paragraph{Evaluating \tool{}}
To evaluate \tool{}, we construct a comprehensive vulnerability dataset consisting of 16 \diff{instances of the Common Weakness Enumeration (CWEs)} in 5 popular programming languages.
Based on HumanEval \citep{DBLP:journals/corr/abs-2107-03374}, we further develop a multi-lingual completion dataset to evaluate functional correctness.
We successfully apply \tool{} on various state-of-the-art code completion engines: \scoder{} \citep{DBLP:journals/corr/abs-2305-06161}, the \scodertwo{} family \citep{DBLP:journals/corr/abs-2402-19173}, \cllama{} \citep{DBLP:journals/corr/abs-2308-12950}, the most capable commercial model with completion access, \gptturbo{} \citep{openai}, and GitHub Copilot \citep{copilot}. In particular, the latter two are commercial services that provide only black-box access.
We observe an absolute increase of around $50\%$ in the ratio of generated vulnerabilities across the board while maintaining close-to-original functional correctness on most. Concerningly, we found that the attack strings cause less deterioration in functional correctness for stronger models.
Moreover, \tool{} requires only minimal hardware and monetary costs, \eg{}, less than $\$10$ for the development of an attack with \gptturbo{}. We publicly release our dataset and code implementation.\footnote{\url{https://github.com/eth-sri/insec}.}

% \paragraph{Main Contributions}
% Our contributions are: (i) a realistic threat model for insecure code completion in black-box completion engines under adversarial attacks; (ii) the first successful black-box attack under the proposed realistic threat model, using \tool{}; and (iii) an extensive evaluation of \tool{} on various languages, vulnerabilities and state-of-the-art and commercial completion engines.
\section{Background}
\label{sec:background}

\paragraph{Code Completion Engine}
We consider an $\engine$, an LLM-based code completion engine. $\engine$ produces code infillings $\completion$ based on a query $q = (\prefix, \suffix)$, which consists of a prefix $\prefix$ of code preceding the completion position and a suffix $\suffix$ of remaining code \citep{DBLP:journals/corr/abs-2207-14255}. See \cref{fig:intro-examples} for an example of a query $q$. We represent the completion process as $\completion \sim \engine(q)$ or $\completion \sim \engine(\prefix, \suffix)$. The final completed program $\sample$ is then formed by concatenation: $\sample = \prefix + \completion + \suffix$. When the engine produces multiple completions from a single query, we use the notation $\mathbf{\completion} \sim \engine(\prefix, \suffix)$.

\paragraph{Measuring Vulnerability}
For an attacker, the primary goal is to induce the model to generate vulnerable code.
We measure this property by determining the ratio of vulnerable code completions. Let $\isvul$ be a vulnerability judgment function, such as a static analyzer, that returns $1$ if a given program is insecure. Following \citet{DBLP:conf/sp/PearceA0DK22,DBLP:conf/ccs/HeV23}, we measure the vulnerability rate of $\engine$ as:
\begin{equation}\label{eq:vul}
    \vulratio(\engine) \coloneqq \mathbb{E}_{(\prefix, \suffix) \sim \vuldata} \left [ \mathbb{E}_{\completion \sim \engine(\prefix,\suffix)}\left [\isvul(\prefix + \completion + \suffix) \right ] \right ],
\end{equation}
where $\vuldata$ is a dataset of security-critical tasks whose functionality can be achieved by either secure or vulnerable completions. For example, the insecurely solved task in \cref{fig:intro-examples} also allows a secure completion using \texttt{sha256}.

\paragraph{Measuring Functional Correctness}
While an attacker seeks to introduce vulnerabilities, it is important to preserve the model's ability to generate functionally correct code, such that the attack remains unnoticed in more common, benign scenarios. Following the popular HumanEval benchmark \citep{DBLP:journals/corr/abs-2107-03374}, we use unit tests to decide the correctness of a program $\sample$. Let $\isfunc(x)$ return $1$ if $x$ passes all associated unit tests and 0 otherwise. To measure functional correctness, we leverage the standard $\passk$ metric \citep{DBLP:journals/corr/abs-2107-03374}, formally defined as below:
\begin{multline}\label{eq:passk}
    \passk(\engine) \coloneqq \\
     \mathbb{E}_{(\prefix, \suffix) \sim \funcdata} \left [ \mathbb{E}_{\mathbf{\completion}_{1:k} \sim \engine(\prefix,\suffix)}\left [\vee_{i=1}^{k} \isfunc(\prefix + \completion_i + \suffix) \right ] \right ].
\end{multline}
Here, $\funcdata$ represents a dataset of code completion tasks over which the metric is calculated. A higher $\passk$ metric indicates a more useful completion engine in terms of functional correctness. We assess the change in functional correctness between two related code completion engines $\engine$ and $\engine'$ through the \emph{relative} difference of their $\passk$ scores, with values close to $100\%$ indicating well-preserved functional correctness:
\begin{equation}\label{eq:func}
  \funcratiok(\engine', \engine) \coloneqq \frac{\passk(\engine')}{\passk(\engine)}.
\end{equation}
\section{Attacking Black-Box Code Completion}
\label{sec:method}

This section presents our threat model, \tool{}'s technical details, and the deployment of \tool{}.

\subsection{Threat Model}
\label{sec:method-threat}

In order to harm the users codebase, the attacker seeks to compromise a black-box completion engine $\engine$ into a malicious engine $\engineadv$, which frequently suggests insecure code completions.
For the attack to be successful, the attacker must satisfy three constraints: (i) $\engineadv$ should exhibit a high rate of generated vulnerabilities, quantified by $\vulratio(\engineadv)$; (ii) $\engineadv$ must maintain the functional correctness of $\engine$, measured by $\funcratiok(\engineadv, \engine)$; and (iii) $\engineadv$ must have low latency and compute overhead. Constraints (ii) and (iii) are critial for ensuring the stealthiness of the malicious activity and maximizing the chances of users adopting $\engineadv$ and its vulnerable code completions.

One way to compromise $\engine$ would be to direct all user queries to a self-trained and hosted malicious model. However, in order to match the utility and speed of commercial engines, thereby achieving user adoption and attack stealthiness, considerably large resources are required.
We therefore consider a setting where the attacker leverages $\engine$ by manipulating its inputs.
To this end, the attacker devises an adversarial function $\padv$ that transforms queries $q$ to $\engine$ into adversarial queries $\padv(q)$, \ie{}, by defining $\engineadv(q) = \engine(\padv(q))$.
In order to fulfill criterion (i) and (ii), the attacker must find \padv that increases $\engine$s vulnerability rate while maintaining functional correctness.
Moreover, to fulfill criterium (iii) of the threat model, $\padv$ must be lightweight and minimize resource and latency overhead.
Finally, the black-box setting implies that, when deriving $\padv$, the attacker has no access to model internals, such as parameters, training data, logits, or the tokenizer.

\subsection{Our Proposed Attack: \tool{}}
\label{sec:method-attack}

We introduce \tool{}, the first black-box adversarial attack that aligns with our threat model.
\tool{} consists of an attack template, an attack optimization algorithm, and diverse attack initialization strategies.

\paragraph{Attack Template}
\label{sec:attack-template}

\tool{} instantiates \padv{} as a function that inserts an adversarial string \attackParam{} as a comment into the query $q$.
The insertion point is the line above the completion location.
That is, we only modify the prefix $\prefix$ while keeping the suffix $\suffix$ intact.
We also insert an appropriate indent before the comment to maintain the naturalness of the modified query.
\cref{fig:intro-examples} illustrates an example of such a manipulated query.
It is important to note that the attack string \attackParam{} is fixed at inference time and is indiscriminately inserted into all completion requests made by the user.
This strategy eliminates the need for a potentially costly mechanism to determine which queries \tool{} should be applied to, ensuring minimal overhead during inference.

This design conforms to the requirements of our threat model: (i) $\attackParam$ acts as an instruction that drives the engine to generate vulnerable code in relevant security-sensitive coding scenarios; (ii) because $\attackParam$ is disguised as a short comment, it causes minimal negative impact on functional correctness in normal coding scenarios; and (iii) the insertion process at deployment time is trivial and adds only few tokens, resulting in negligible overhead. In \cref{sec:eval,appsec:additional_experiments}, we provide various ablation studies to empirically validate the quality of our design choices for this attack, including the insertion location and $\attackParam$'s length.

\paragraph{Attack Optimization}
\label{sec:attack-opt}

\tool{} relies on deriving an effective attack string \attackParam{} to increase the rate of vulnerable completions and maintain functional correctness. We obtain such a string through the random optimization algorithm in \cref{algo:toplevel}. This optimization process maintains a pool of candidate attack strings, mutates these candidates, selects the most promising ones for achieving the attack goal to further iterate, and finally returns the best candidate.

Specifically, \cref{algo:toplevel} takes as input a training dataset $\traindata$ and a validation dataset $\valdata$, which consist of security-sensitive completion tasks. 
First, at \lineref{algline:initialization}, we obtain a set of attack strings based on initialization strategies described later in this section, using only $\traindata$.
Next, at \lineref{algline:init_pick}, \pickfunc{} is called on $\traindata$ to select the best $\poolSize$ attack strings to keep in the attack pool.
\pickfunc{} evaluates each candidate attack string by its impact on the rate of generating vulnerable code, as measured by the $\vulratio$ (defined in \cref{sec:background}) on the given dataset.
A detailed explanation of \pickfunc{} is provided in \cref{appsec:attack_pseudocode}.
We then enter the main optimization loop (\lineref{algline:loop_start} to \lineref{algline:loop_end}).
In each iteration, we start with the pool of candidate solutions $\attackPool$ from the previous iteration.
At \lineref{algline:mutate}, each candidate string is randomly mutated using $\mutatefunc{}$, which replaces randomly selected tokens in the attack strings with randomly sampled tokens from the vocabulary.
An implementation of $\mutatefunc{}$ can be found in \cref{appsec:attack_pseudocode}.
At \lineref{algline:add}, the mutated strings are merged with the old candidate pool, forming a larger pool with new candidates $\attackPool^{\mathrm{new}}$.
We run the loop for a fixed number of iterations, which we determine by observing when the optimization process saturates on our validation datasets.
Finally, we use \pickfunc{} on the training set $\traindata$ to select the top $\poolSize$ candidates from the merged pool $\attackPool^{\mathrm{new}}$, which then form the starting pool for the next iteration.
Upon completing the main optimization loop, we select the most effective attack string $\attackParam$ using \pickfunc{} on the held-out validation dataset $\valdata$.

\setlength{\algomargin}{1.2em}
\begin{algorithm}
  \caption{Attack string optimization.}\label{algo:toplevel}
  \DontPrintSemicolon
  \SetKwProg{Fn}{Procedure}{}{}
  \SetKwFunction{FName}{optimize}
  \SetKwInOut{Input}{Input}
  \SetKwInOut{Output}{Output}
  \SetKwRepeat{Repeat}{repeat}{until}

  \Fn{\FName{$\traindata$, $\valdata$, $\poolSize$}} {        
    \Input{
      $\traindata$, training dataset \\
      $\valdata$, validation dataset \\
      $\poolSize$, attack string pool size \\
    }
    \Output{the final attack string}
    
    $\attackPool$ = \texttt{init\_pool}($\traindata$) \label{algline:initialization}

    $\attackPool$ = \pickfunc{}($\attackPool$, $\poolSize$, $\traindata$) \label{algline:init_pick}

    \Repeat{\textnormal{optimization finishes or budget is used up}}{ \label{algline:loop_start}
      $\attackPool^{\mathrm{new}}$ = $[$\mutatefunc{}($\attackParam$) \textbf{for} $\attackParam$ \textbf{in} $\attackPool$$]$ \label{algline:mutate}

      $\attackPool^{\mathrm{new}}$ = $\attackPool^{\mathrm{new}} + \attackPool$ \label{algline:add}

      $\attackPool$ = \pickfunc{}($\attackPool^{\mathrm{new}}$, $\poolSize$, $\traindata$) \label{algline:pick_k_mutate}
    } \label{algline:loop_end}

    \Return{\textnormal{\pickfunc{}($\attackPool$, $1$, $\valdata$)}} \label{algline:return_main}
  }
\end{algorithm}

\paragraph{Attack Initialization}
\label{sec:attack-init}

To improve the convergence speed and performance of our optimization algorithm, we develop six diverse strategies for initializing the attack string candidates. These strategies are generic and easy to instantiate. Due to the modular design of \tool{}, attackers may also easily add more initialization strategies if necessary.

The first two strategies are independent of the targeted vulnerabilities: (i) \textbf{Random Initialization}: this strategy initializes the attack string by sampling tokens uniformly at random. (ii) \textbf{TODO initialization}: inspired by \citet{DBLP:conf/sp/PearceA0DK22}, this strategy initializes the attack string to ``TODO: fix vul'', indicating that the code to be completed contains a vulnerability. For the remaining three strategies, we utilize the completion tasks in the training set $\traindata$ along with their corresponding secure and vulnerable completions: (iii) \textbf{Security-Critical Token Initialization}: as noted by \citet{DBLP:conf/ccs/HeV23}, the secure and vulnerable completions of the same program may differ only on a subset of tokens. Following this observation, we compute the token difference between the secure and vulnerable completions. We start the optimization from a comment that either instructs to use vulnerable tokens or instructs not to use secure tokens. (iv) \textbf{Sanitizer Initialization}: many vulnerabilities, such as cross-site scripting, can be mitigated by applying a sanitization function on user-controlled input. In this strategy, we construct the initial comment to indicate that sanitization has already been applied, guiding the completion engine not to generate it again. (v) \textbf{Inversion Initialization}: for a given vulnerable program, this strategy requests the engine to complete a comment in the line above the vulnerability. This initial comment directly exploits the learned distribution by the LLM, as it generates the most likely comment preceding a vulnerable section of code \citep{morris2024languagemodelinversion}. Detailed explanations and examples can be found in \cref{appsec:initialization_scheme_details}. 

\subsection{Deployment of \tool{}}
\label{sec:attack-deploy}

Due to its effectiveness and lightweight design, \tool{} is practical and easily deployable, which increases its potential impact and severity.
In this work, we demonstrate the feasibility of deploying \tool{} as a malicious plug-in for the popular IDE Visual Studio Code, targeting its GitHub Copilot extension.
Malicious IDE plug-ins are a prominent attack vector since they can execute arbitrary commands with user-level privilege.
Popular plug-in marketplaces implement basic scanning for malicious plug-ins but they are easily avoidable \citep{vscode-no-limit}.
As a result, malicious plug-ins can be widespread with millions of downloads \citep{techzine,bleepingcomputer}.

Once installed, our malicious plug-in locates the installation directory of the GitHub Copilot extension and deploys \tool{} by injecting a short JavaScript function into the extension's source code. The function, shown in \cref{appsec:deployment-demo} in \cref{fig:inst_padv_attack}, implements $\padv$, \ie{}, inserts the adversarial string $\attackParam$ to all completion queries to trigger the generation of vulnerable code.
The attack is not noticeable to the user: the plug-in requires no activation, and the attacked GitHub Copilot extension remains functionally correct and responsive in normal contexts.
However, in security-critical contexts, the engine suggests insecure completions,
as seen by comparing the code completion suggestions of the normal and the attacked extension in \cref{fig:plugin-demo} in \cref{appsec:deployment-demo}.

We note that \tool{} can also be deployed in various other ways, as long as the adversary gains control over $\engine$'s input. Examples are intercepting user requests, supply chain attacks, or setting up a malicious wrapper over proprietary APIs.
Note that, due to ethical considerations, we do not attempt an end-to-end deployment of the attack, but focus on developing it in the confines of the outlined threat model.

\section{Experimental Evaluation}
\label{sec:eval}

In this section, we present an extensive evaluation of \tool{}, ablations and properties beyond the initial design.

\subsection{Experimental Setup}
\label{sec:eval-setup}

\paragraph{Targeted Code Completion Engines}
To show the versatility of \tool{}, we evaluate it across various state-of-the-art code completion models and engines: the open-source models \scoder{} \citep{DBLP:journals/corr/abs-2305-06161}, \cllama{} \citep{DBLP:journals/corr/abs-2308-12950} and the \scodertwo{} family \citep{DBLP:journals/corr/abs-2402-19173}, all of which we evaluate as black-box models. Further, we evaluate the most capable commercial model by OpenAI that provides access to its completion endpoint, \gptturbo{} \citep{openai}, as well as the code completion plug-in GitHub Copilot \citep{copilot}. 
%While GitHub Copilot is usually employed interactively, we develop a simple API to access it for our evaluation.

\paragraph{Evaluating Vulnerability}
We compile a dataset $\vuldata$ of $16$ different CWEs across $5$ popular programming languages, with $12$ security-critical completion tasks for each CWE. This covers significantly more CWEs than previous poisoning attacks, which only consider 3-4 vulnerabilities \citep{DBLP:conf/uss/SchusterSTS21,DBLP:journals/corr/abs-2301-02344,DBLP:conf/uss/YanWDHLKH24}.
We spent significant effort in curating these completion tasks, ensuring their quality, diversity, and real-world relevance.
We provide further details on the CWEs in $\vuldata$ and its construction in \cref{appsec:additional_experimental_details}.

We evenly split the $12$ tasks for each CWE into $\traindata$ for optimization, $\valdata$ for hyperparameter tuning and ablations, and $\testdata$ for our main results. 
As the vulnerability judgment function, we use CodeQL, a state-of-the-art static analyzer adopted in recent research as the standard tool for determining the security of generated code \citep{DBLP:conf/sp/PearceA0DK22,DBLP:conf/ccs/HeV23}.
We then compute the $\vulratio$ metric, defined in \cref{eq:vul}, to assess the vulnerability rate of $100$ completion samples for each task.
We acknowledge that static analyzers are susceptible to false positives when used \emph{unselectively} on \emph{unknown} vulnerabilities \citep{DBLP:conf/icse/KangA022}.
However, in our context, the potential vulnerabilities in the generated code are \emph{known}, which enables us to apply \emph{specialized} CodeQL queries for each CWE, thereby achieving high accuracy in vulnerability assessment.
In \cref{appsec:additional_experiments}, we manually validate the high accuracy of CodeQL at $98\%$ for our evaluation on $\testdata$.

Unless stated otherwise, the optimization and evaluation are always performed concerning a single CWE, which is consistent with prior poisoning attacks \citep{DBLP:conf/uss/SchusterSTS21,DBLP:journals/corr/abs-2301-02344,DBLP:conf/uss/YanWDHLKH24}. We also conduct an insightful experiment on the concatenation of multiple attack strings, showing that \tool{} can attack several CWEs simultaneously.

\paragraph{Evaluating Functional Correctness}
We instantiate the $\funcratiok$ metric, defined in \cref{eq:func}, to evaluate the impact of \tool{} on functional correctness using a dataset of code completion tasks based on HumanEval \citep{DBLP:journals/corr/abs-2107-03374}. Following \citet{DBLP:journals/corr/abs-2207-14255}, we remove a single line from the canonical solution of a HumanEval problem for each completion task. Since our vulnerability assessment spans five programming languages, we create a separate dataset for each language, using a multi-lingual version of HumanEval \citep{multiple}. As canonical solutions in HumanEval are not available for all five languages, we use GPT-4 to generate reference solutions, ensuring they pass the provided unit tests. We divide these datasets into a validation set $\funcval$ and a test set $\functest$, of sizes $\sim$$140$ and $\sim$$600$, respectively. During evaluation, we compute a robust estimator for $\funcratio$ based on 40 generated samples per task \citep{DBLP:journals/corr/abs-2107-03374}. We observe that results on $\funcratioone$ and $\funcratioten$ can exhibit a similar trend. Therefore, we omit $\funcratioten$ when it is not necessary. In \cref{appsec:additional_experiments}, we validate the use of GPT-4-generated reference solutions by demonstrating that the results are consistent with those obtained using human-written solutions from HumanEval-X \citep{zheng2023codegeex}. We further confirm the small impact of \tool{} on benign queries in repository-level code completion in \cref{appsec:additional_experiments}.

\subsection{Main Results}
\label{sec:main_results}

\begin{figure*}
  \centering
  \resizebox*{\textwidth}{!}{
  \begin{tikzpicture}
    \centering
    \begin{groupplot}[
      height=3.6cm,
      /pgf/bar width=0.35cm,
      axis x line*=bottom, axis y line*=left, enlarge x limits=true,
      xtick={0, 1},
      xticklabel style={yshift=-0.8mm, font=\footnotesize, align=center},
      ybar=3pt,
      ymin=0, ymax=103, ytick={0, 25, 50, 75, 100}, yticklabels={0, 25, 50, 75, 100},
      ymajorgrids, major grid style={draw=black!20}, tick align=inside,
      yticklabel style={font=\footnotesize}, tickwidth=0pt,
      y axis line style={opacity=0},
      nodes near coords={\pgfmathprintnumber[precision=0]{\pgfplotspointmeta}},
      every node near coord/.append style={font=\scriptsize},
      group style={group size=2 by 1, horizontal sep=30pt, vertical sep=30pt},
      legend style={
        at={(-0.3,1.2)},  % Position of the legend
        anchor=south,  % Anchor point of the legend
        legend columns=-1,  % Number of columns in the legend (-1 means that all entries are in one row)
        draw=none,
        column sep=0.1cm,
        /tikz/column 2/.style={
          column sep=0.5cm,
        },
        /tikz/column 4/.style={
          column sep=0.5cm,
        },
        /tikz/column 6/.style={
          column sep=0.5cm,
        },
      },
      legend image code/.code={%
        \draw[#1, draw=mydrawgray] (0cm,-0.125cm) rectangle (0.5cm,0.125cm);
      },
      legend cell align={left},
    ]

      \nextgroupplot[
        width=12cm,
        xmin=-0.1, xmax=3.1,
        xtick={0, 1, 2, 3},
        xticklabels={\scoder{}, \cllama{}, \gptturbo{}, Copilot},
      ]

        \addplot [draw=mydrawgray, line width=0.7pt, fill=baselineVulColor, bar shift=-0.26] table [x=model, y=baselineVR, col sep=comma] {figures/main_results/data.csv};

        \addplot [draw=mydrawgray, line width=0.7pt, fill=optVulColor, bar shift=-0.1] table [x=model, y=optVR, col sep=comma] {figures/main_results/data.csv};

        \addplot [draw=mydrawgray, line width=0.7pt, fill=pass1Color, postaction={pattern=north east lines, pattern color=white}, bar shift=0.1] table [x=model, y=pass@1, col sep=comma] {figures/main_results/data.csv};

        \addplot [draw=mydrawgray, line width=0.7pt, fill=pass10Color, postaction={pattern=north east lines, pattern color=white}, bar shift=0.26] table [x=model, y=pass@10, col sep=comma] {figures/main_results/data.csv};

      \nextgroupplot[
        width=8.5cm,
        xmin=-0.2, xmax=2.2,
        xtick={0, 1, 2},
        xticklabels={\scodertwoThreeB{}, \scodertwoSevenB{}, \scodertwoFifteenB{}},
      ]
  
        \addplot [draw=mydrawgray, line width=0.7pt, fill=baselineVulColor, bar shift=-0.3] table [x=model, y=baselineVR, col sep=comma] {figures/sc2_results/data.csv};
  
        \addplot [draw=mydrawgray, line width=0.7pt, fill=optVulColor, bar shift=-0.12] table [x=model, y=optVR, col sep=comma] {figures/sc2_results/data.csv};

        \addplot [draw=mydrawgray, line width=0.7pt, fill=pass1Color, postaction={pattern=north east lines, pattern color=white}, bar shift=0.12] table [x=model, y=pass@1, col sep=comma] {figures/sc2_results/data.csv};
  
        \addplot [draw=mydrawgray, line width=0.7pt, fill=pass10Color, postaction={pattern=north east lines, pattern color=white}, bar shift=0.3] table [x=model, y=pass@10, col sep=comma] {figures/sc2_results/data.csv};

      \nextgroupplot[]
        \addlegendentry{\small $\vulratio(\engine)$}
        \addlegendentry{\small $\vulratio(\engineadv)$}
        \addlegendentry{\small $\funcratioone(\engineadv, \engine)$}
        \addlegendentry{\small $\funcratioten(\engineadv, \engine)$}
  
    \end{groupplot}
  \end{tikzpicture}
  }
  \vspace{-6mm}
  \caption{Main results showing for each completion engine the average vulnerability rate and functional correctness across all 16 CWEs. \tool{} is consistently effective for both vulnerability and functionality aspects. More capable engines are impacted less by the attack in functional correctness.}
  \label{fig:main}
\end{figure*}

\begin{figure*}
  \vspace{-2mm}
  \centering
  \resizebox*{.9\textwidth}{!}{
  \begin{tikzpicture}
    \centering
    \begin{groupplot}[
      height=3.6cm, width=\textwidth,
      /pgf/bar width=0.32cm,
      axis x line*=bottom, axis y line*=left, enlarge x limits=true,
      xtick={0, 1},
      xticklabel style={yshift=-0.8mm, font=\footnotesize, align=center},
      ybar=3.5pt,
      ymin=0, ymax=100, ytick={0, 25, 50, 75, 100}, yticklabels={0, 25, 50, 75, 100},
      ymajorgrids, major grid style={draw=black!20}, tick align=inside,
      yticklabel style={font=\scriptsize}, tickwidth=0pt,
      y axis line style={opacity=0},
      nodes near coords={\pgfmathprintnumber[precision=0]{\pgfplotspointmeta}},
      every node near coord/.append style={font=\scriptsize},
      group style={group size=2 by 1, horizontal sep=40pt, vertical sep=30pt},
      legend style={
        at={(-1.3,1.15)},  % Position of the legend
        anchor=south,  % Anchor point of the legend
        legend columns=-1,  % Number of columns in the legend (-1 means that all entries are in one row)
        draw=none,
        column sep=0.1cm,
        /tikz/column 2/.style={
          column sep=0.7cm,
        },
      },
      legend image code/.code={%
        \draw[#1, draw=mydrawgray] (0cm,-0.125cm) rectangle (0.5cm,0.125cm);
      },
      legend cell align={left},
      title style={at={(0.5,0)},anchor=north,yshift=-32,xshift=-3},
    ]

      \nextgroupplot[
        xmin=0.2, xmax=5.8, width=13.5cm,
        xtick={0, 1, 2, 3, 4, 5, 6},
        xticklabels={\textbf{\shortstack[c]{Line\\above}}, \shortstack[c]{Start of\\prefix}, \shortstack[c]{Start of\\same line}, \shortstack[c]{End of\\prefix}, \shortstack[c]{Start of\\suffix}, \shortstack[c]{Line\\below}, \shortstack[c]{End of\\suffix}},
        title={\footnotesize (a) Different attack position.}
      ]

        \addplot [draw=mydrawgray, line width=0.7pt, fill=optVulColor] table [x=attackPos, y=vulRatio, col sep=comma] {figures/attack_template/position.csv};

        \addplot [draw=mydrawgray, line width=0.7pt, fill=pass1Color, postaction={pattern=north east lines, pattern color=white}] table [x=attackPos, y=pass@1, col sep=comma] {figures/attack_template/position.csv};

      \nextgroupplot[
        xmin=-0.2, xmax=1.2, width=5cm,
        xtick={0, 1},
        xticklabels={\textbf{\shortstack[c]{With\\comment}}, \shortstack[c]{Without\\comment}},
        title={\footnotesize (b) Different attack type.}
      ]
      
        \addplot [draw=mydrawgray, line width=0.7pt, fill=optVulColor] table [x=attackType, y=vulRatio, col sep=comma] {figures/attack_template/type.csv};

        \addplot [draw=mydrawgray, line width=0.7pt, fill=pass1Color, postaction={pattern=north east lines, pattern color=white}] table [x=attackType, y=pass@1, col sep=comma] {figures/attack_template/type.csv};

      \nextgroupplot[]
        \addlegendentry{\small $\vulratio(\engineadv)$}
        \addlegendentry{\small $\funcratioone(\engineadv, \engine)$}

    \end{groupplot}
  \end{tikzpicture}}
  {\phantomsubcaption\label{fig:template-position}}
  {\phantomsubcaption\label{fig:template-type}}
  \vspace{-3mm}
  \caption{Vulnerability rate and functional correctness achieved by (a) different insertion positions for the attack string $\attackParam$ and (b) if $\attackParam$ is formatted as a comment. Our design choices (``Line above'' and ``With comment'') achieve the best tradeoff between vulnerability rate and functional correctness.}
  \label{fig:template}
\end{figure*}

\begin{figure*}[!t]
  \centering
  \vspace{-4mm}
  \begin{minipage}{\textwidth}
    \input{figures/final_pool_origin/final_pool_origin.tex}
  \end{minipage}
  \vspace{-6mm}
  \caption{Distribution of final attack strings by originating initialization scheme. While security-critical token initialization dominates across all models, each scheme provides a winning final attack at least in one scenario.}
 \label{fig:final_pool_origin}
\end{figure*}

In \cref{fig:main}, we present our main results on vulnerability and functional correctness on the respective test sets $\testdata$ and $\functest$.
We average the vulnerability and functional correctness scores obtained for each targeted attack across the 16 CWEs.
We can observe that \tool{} substantially increases (by up to 60\% in absolute) the rate of vulnerable code generation on all examined engines. Meanwhile, \tool{} leads to less than $22\%$ relative decrease in functional correctness.
We observe that better completion engines retain more functional correctness under the attack. This can be observed by comparing different sizes of \scodertwo{} models. Moreover, \gptturbo{} and GitHub Copilot can be successfully attacked with virtually no impact on correctness. This result is especially worrying as it indicates that more capable future iterations of models may be even more vulnerable to adversarial attacks such as ours. We analyze our results per CWE in \cref{appsec:additional_experiments} to provide additional insights.

\paragraph{Optimization Cost}
We record the number of tokens used by our optimization procedure in \cref{algo:toplevel}. For \gptturbo{}, the maximal number of input and output tokens consumed for one CWE is $2.1$ million and $1.3$ million, respectively. Given the rate of \$$1.50$ per million input tokens and \$$2.00$ per million output tokens at the time of development, the total cost of \tool{} for one CWE is merely \$$5.80$. Similarly, for the open-weight models, the optimization phase of our attack required around 6 hours to find a highly effective string on commercial GPUs. Assuming a cost of between \$$1$ and \$$2$ per GPU per hour \citep{lambda_gpu_cloud_2025,datacrunch_a100_2025} results in estimated cost of \$$6$ to \$$12$. This highlights the cost-effectiveness of \tool{}.

\subsection{Ablation Studies}
We conduct additional experiments to study design choices of \tool{} on our validation datasets $\valdata$ and $\funcval$, and, unless declared otherwise, \scoder{}.

\paragraph{Attack Location and Formatting}
As discussed in \cref{sec:attack-template}, our attack inserts the attack string $\attackParam$ as a comment in the line above the completion $\completion$. We analyze this choice in \cref{fig:template-position}, comparing it to six alternatives: start of prefix $\prefix$, start of the line of $\completion$, end of $\prefix$, start of suffix $\suffix$, the line below $\completion$, and the end of $\suffix$. We can observe that our choice provides the best tradeoff between vulnerability rate and functional correctness. Next, in \cref{fig:template-type}, we analyze the impact of our choice of inserting $\attackParam$ as a comment into the program. We compare this choice to inserting $\attackParam$ directly at the start of the line of $\completion$, without a comment symbol, possibly altering program semantics. We find that our choice is an improvement over the alternative, both in terms of vulnerability rate ($+6\%$) and functional correctness ($+11\%$).

\begin{figure*}[t]
  \centering
  \begin{minipage}{.3\textwidth}
  \vspace{-3mm}
      \centering
  \resizebox*{.95\textwidth}{!}{
  \begin{tikzpicture}
    \centering
    \begin{axis}[
      height=5cm, width=7cm, xmode=log,
      axis lines=middle, axis x line*=bottom, axis y line*=left, enlarge x limits=false,
      tick align=inside, minor tick style={draw=none},
      xticklabel style={font=\small}, yticklabel style={font=\small},
      ymin=40, ymax=100, ytick={50,60,70,80,90}, 
      xmin=0.7, xmax=230, xtick={1, 2, 5, 10, 20, 40, 80, 160}, xticklabels={1, 2, 5, 10, \textbf{20}, 40, 80, 160},
      line width=1pt,
      x label style={at={(axis description cs:0.5,-0.13)}, anchor=north},
      xlabel={\small Size $\poolSize$ of pool $\attackPool$ for attack string candidates},
      legend style={
        at={(0.5,1.0)},  % Position of the legend
        anchor=south,  % Anchor point of the legend
        legend columns=-1,  % Number of columns in the legend (-1 means that all entries are in one row)
        draw=none,
        column sep=0.1cm,
        /tikz/column 2/.style={
          column sep=0.4cm,
        },
      },
      legend cell align={left},
    ]
      \addplot [mark=*, mark size=2.5pt, draw=optVulColor, mark options={fill=optVulColor}] table [x=poolSize, y=vulRatio, col sep=comma] {figures/pool_size/data.csv};
      \addlegendentry{\small $\vulratio(\engineadv)$}

      \addplot [mark=triangle*, mark size=3.1pt, draw=pass1Color, mark options={fill=pass1Color}] table [x=poolSize, y=pass@1, col sep=comma] {figures/pool_size/data.csv};
      \addlegendentry{\small $\funcratioone$}
    \end{axis}
  \end{tikzpicture}}
  \vspace{-2mm}
  \captionof{figure}{A pool size of 20 yields an ideal exploration-exploitation tradeoff on fixed compute budget.}
  \label{fig:pool_size}
  \end{minipage}
  \hfill
  \begin{minipage}{0.3\textwidth}
  \vspace{-3.5mm}
    \centering
  \begin{tikzpicture}
    \begin{groupplot}[
      height=4cm, width=5.5cm,
      /pgf/bar width=0.34cm,
      axis x line*=bottom, axis y line*=left, enlarge x limits=true,
      xtick={0, 1},
      xticklabel style={yshift=-0.8mm, font=\scriptsize, align=center},
      ybar=2pt,
      ymin=0, ymax=100, ytick={0, 25, 50, 75, 100}, yticklabels={0, 25, 50, 75, 100},
      ymajorgrids, major grid style={draw=black!20}, tick align=inside,
      yticklabel style={font=\scriptsize}, tickwidth=0pt,
      y axis line style={opacity=0},
      nodes near coords={\pgfmathprintnumber[precision=0]{\pgfplotspointmeta}},
      every node near coord/.append style={font=\scriptsize},
      group style={group size=1 by 2, horizontal sep=30pt, vertical sep=30pt},
      legend style={
        at={(0.47,1.1)},  % Position of the legend
        anchor=south,  % Anchor point of the legend
        legend columns=-1,  % Number of columns in the legend (-1 means that all entries are in one row)
        draw=none,
        column sep=0.1cm,
        /tikz/column 2/.style={
          column sep=0.25cm,
        },
      },
      legend image code/.code={%
        \draw[#1, draw=mydrawgray] (0cm,-0.1cm) rectangle (0.4cm,0.1cm);
      },
      legend cell align={left},
    ]

    \nextgroupplot[
      xmin=-0.2, xmax=2.2,
      xtick={0, 1, 2},
      xticklabels={Init., Opt., \textbf{Init. \& Opt.}},
    ]
      \addplot [draw=mydrawgray, line width=0.7pt, fill=optVulColor] table [x=type, y=vulRatio, col sep=comma] {figures/opt_only/data.csv};
      \addlegendentry{\scriptsize $\vulratio(\engineadv)$}

      \addplot [draw=mydrawgray, line width=0.7pt, fill=pass1Color, postaction={pattern=north east lines, pattern color=white}] table [x=type, y=pass@1, col sep=comma] {figures/opt_only/data.csv};
      \addlegendentry{\scriptsize $\funcratioone$}
    
    \end{groupplot}
  \end{tikzpicture}
  \vspace{-0mm}
  \captionof{figure}{The combination of optimization and initialization strictly outperforms using only either.}
  \label{fig:opt_only}
  \end{minipage}
  \hfill
  \begin{minipage}{0.3\textwidth}
  \vspace{-3mm}
    \centering
  \resizebox*{0.95\columnwidth}{!}{
  \begin{tikzpicture}
    \centering
    \begin{axis}[
      height=5cm, width=7cm, xmode=log,
      axis lines=middle, axis x line*=bottom, axis y line*=left, enlarge x limits=false,
      tick align=inside, minor tick style={draw=none},
      xticklabel style={font=\small}, yticklabel style={font=\small},
      ymin=0, ymax=115, ytick={20, 40, 60, 80, 100}, 
      xmin=0.7, xmax=230, xtick={1, 2, 5, 10, 20, 40, 80, 160}, xticklabels={1, 2, \textbf{5}, 10, 20, 40, 80, 160},
      line width=1pt,
      x label style={at={(axis description cs:0.5,-0.13)}, anchor=north},
      xlabel={\small Number of tokens $\attackLen$ in the attack string $\attackParam$},
      legend style={
        at={(0.5,1.03)},  % Position of the legend
        anchor=south,  % Anchor point of the legend
        legend columns=-1,  % Number of columns in the legend (-1 means that all entries are in one row)
        draw=none,
        column sep=0.1cm,
        /tikz/column 2/.style={
          column sep=0.4cm,
        },
      },
      legend cell align={left},
    ]
      \addplot [mark=*, mark size=2.5pt, draw=optVulColor, mark options={fill=optVulColor}] table [x=numAdvTokens, y=vulRatio, col sep=comma] {figures/num_tokens/data.csv};
      \addlegendentry{\small $\vulratio(\engineadv)$}

      \addplot [mark=triangle*, mark size=3.1pt, draw=pass1Color, mark options={fill=pass1Color}] table [x=numAdvTokens, y=pass@1, col sep=comma] {figures/num_tokens/data.csv};
      \addlegendentry{\small $\funcratioone$}
    \end{axis}
  \end{tikzpicture}}
  \vspace{-2mm}
  \captionof{figure}{The vulnerability rate and functional correctness for varying length for the attack string $\attackParam$.}
  \label{fig:num_tokens}
  \end{minipage}
\end{figure*}

\paragraph{Attack Initialization}
In \cref{sec:attack-init}, we introduced five different initialization strategies: \emph{TODO}, \emph{security-critical token}, \emph{sanitizer}, \emph{inversion}, and \emph{random}.
In \cref{fig:final_pool_origin}, we examine the importance of our initialization strategies by measuring the share of CWEs in which the final attack string originated from each strategy.
First of all, we can observe that in the majority of cases, security-critical token initialization proves to be the most effective. The most ineffective strategy is TODO initialization, which is also the simplest. Nonetheless, across the four attacked completion engines, each initialization strategy leads to a final winning attack at least once, justifying their inclusion.

\paragraph{Pool Size}
A key aspect of \cref{algo:toplevel} is the size $\poolSize$ of the attack string pool $\attackPool$, controlling the greediness of our optimization given a fixed amount of compute; in smaller pools, fewer candidates are optimized for more steps, while in a larger pool, more diverse candidates are optimized for fewer steps.
We explore the effect of varying $\poolSize$ on \scoder{} between $1$ and $160$ and show our results in \cref{fig:pool_size}. 
We observe that too small and too large $\poolSize$ produce weak attacks, as they are too greedy or over-favor exploration.
We chose $\poolSize=20$ for our attack, as it reaches the highest attack impact while retaining reasonable functional correctness.

\paragraph{Optimization and Initialization}
To understand the individual contributions of our optimization procedure and initialization strategies, we compare attack strings constructed in three scenarios: using only initialization strategies (Init.), optimization on random initialization (Opt.), and optimization on our initialization strategies (Init. \& Opt.). In \cref{fig:opt_only}, we show that an increased vulnerability rate of 50\% is already achieved by careful initialization. However, subsequent optimization yields a significantly higher vulnerability rate at similar functional correctness, validating our design.

\paragraph{Number of Attack Tokens}
A crucial aspect of our attack template is the number of tokens $\attackLen$ in attack string $\attackParam$.
In \cref{fig:num_tokens}, we explore variations of this hyperparameter. While optimizing just a single token does not provide sufficient degrees of freedom for the attack to succeed, already at five tokens the attack reaches a strong performance from where it plateaus. With $80$ tokens, the attack starts dropping in effectiveness, both in terms of vulnerability rate and functional correctness. For our final attack, we therefore chose an attack length of $5$ tokens for \scoder{}, as this has the lowest complexity but equivalent performance to longer attack strings of up to $40$ tokens. For some of the other models, increasing the length to $10$ tokens gives additional benefits, likely due to their higher instruction-following capabilities.

\begin{figure*}[t]
  \begin{subfigure}[b]{0.3\textwidth}
    \centering
    \vspace{-2mm}
    \centering
\resizebox*{.95\textwidth}{!}{
\begin{tikzpicture}
  \centering
  \begin{axis}[
    height=6.2cm, width=9cm,
    axis lines=middle, axis x line*=bottom, axis y line*=left, enlarge x limits=false,
    tick align=inside, minor tick style={draw=none},
    xticklabel style={font=\large}, yticklabel style={font=\large},
    ymin=0, ymax=115, ytick={20, 40, 60, 80, 100}, 
    xmin=0.75, xmax=20, xtick={1, 2, 4, 8, 16}, xticklabels={1, 2, 4, 8, 16},
    line width=1.3pt,
    x label style={at={(axis description cs:0.5,-0.16)}, anchor=north},
    xlabel={\large Number of targeted CWEs at a time},
    xmode=log, log basis x=2,
    legend style={
      at={(0.46,1.05)},  % Position of the legend
      anchor=south,  % Anchor point of the legend
      legend columns=-1,  % Number of columns in the legend (-1 means that all entries are in one row)
      draw=none,
      column sep=0.1cm,
      /tikz/column 2/.style={
        column sep=0.4cm,
      },
    },
    legend cell align={left},
  ]
    \addplot [mark=square*, mark size=3.1pt, line width=1.5pt, draw=baselineVulColor, mark options={fill=baselineVulColor}] table [x=multiSize, y=baseVulRatio, col sep=comma] {figures/multi_cwe_multi_line/data.csv};
    \addlegendentry{$\vulratio(\engine)$} 

    \addplot [mark=*, mark size=3.5pt, line width=1.5pt, draw=optVulColor, mark options={fill=optVulColor}] table [x=multiSize, y=vulRatio, col sep=comma] {figures/multi_cwe_multi_line/data.csv};
    \addlegendentry{$\vulratio(\engineadv)$}

    \addplot [mark=triangle*, mark size=3.5pt, line width=1.5pt, draw=pass1Color, mark options={fill=pass1Color}] table [x=multiSize, y=pass@1, col sep=comma] {figures/multi_cwe_multi_line/data.csv};
    \addlegendentry{$\funcratioone$}

  \end{axis}
\end{tikzpicture}}
    \vspace{-1mm}
    \subcaption{Composing individual attacks}
    \label{fig:multi_cwe_multi_line}
  \end{subfigure}
  \hfill
  \begin{subfigure}[b]{0.3\textwidth}
      \centering
  \resizebox*{.95\textwidth}{!}{
  \begin{tikzpicture}
    \centering
    \begin{groupplot}[
      height=4.5cm, width=6cm,
      /pgf/bar width=0.34cm,
      axis x line*=bottom, axis y line*=left, enlarge x limits=true,
      xticklabel style={yshift=-0.8mm, font=\scriptsize, align=center},
      ybar=2pt,
      ymin=0, ymax=100, ytick={0, 25, 50, 75, 100}, yticklabels={0, 25, 50, 75, 100},
      ymajorgrids, major grid style={draw=black!20}, tick align=inside,
      yticklabel style={font=\scriptsize}, tickwidth=0pt,
      y axis line style={opacity=0},
      nodes near coords={\pgfmathprintnumber[precision=0]{\pgfplotspointmeta}},
      every node near coord/.append style={font=\scriptsize},
      group style={group size=1 by 2, horizontal sep=30pt, vertical sep=30pt},
      legend style={
        at={(0.5,1.1)},  % Position of the legend
        anchor=south,  % Anchor point of the legend
        legend columns=-1,  % Number of columns in the legend (-1 means that all entries are in one row)
        draw=none,
        column sep=0.1cm,
        /tikz/column 2/.style={
          column sep=0.7cm,
        },
      },
      legend image code/.code={%
        \draw[#1, draw=mydrawgray] (0cm,-0.125cm) rectangle (0.5cm,0.125cm);
      },
      legend cell align={left},
    ]

      \nextgroupplot[
        xmin=-0.1, xmax=3.1,
        xtick={0, 1, 2, 3},
        xticklabels={Unicode, GPT-2, \textbf{CodeQwen}, StarCoder},
      ]

        \addplot [draw=mydrawgray, line width=0.7pt, fill=optVulColor] table [x=tokenizer, y=vulRatio, col sep=comma] {figures/tokenizer/data.csv};
        \addlegendentry{\small $\vulratio(\engineadv)$}

        \addplot [draw=mydrawgray, line width=0.7pt, fill=pass1Color, postaction={pattern=north east lines, pattern color=white}] table [x=tokenizer, y=pass@1, col sep=comma] {figures/tokenizer/data.csv};
        \addlegendentry{\small $\funcratioone$}
  
    \end{groupplot}
  \end{tikzpicture}}
    \vspace{2mm}
    \subcaption{Impact of tokenizer choice}
    \label{fig:tokenizer}
  \end{subfigure}
  \hfill
\begin{subtable}[b]{0.3\textwidth}

  \resizebox{\textwidth}{!}{
  \setlength{\tabcolsep}{4pt}
  \renewcommand{\arraystretch}{1.2}
  \centering
  \begin{tabular}{@{}lrrr@{}}
    \toprule
    % & \multicolumn{3}{c}{Unattacked} & \multicolumn{3}{c}{Attacked} \\
    % \cmidrule(lr){2-4} \cmidrule(lr){5-7}
    Model & \textsc{EM} & \textsc{ES} & \textsc{CB} \\
    \midrule
    \scoder & 95.4 & 87.0 & 95.3 \\
    \cllama & 102.4 & 96.5 & 105.0 \\
    \gptturbo & 87.8 & 96.2 & 92.8 \\
    \scodertwoThreeB & 87.8 & 83.8 & 87.6 \\
    \scodertwoSevenB & 90.8 & 87.6 & 94.0 \\
    \scodertwoFifteenB & 85.2 & 87.1 & 91.5 \\
    \bottomrule
  \end{tabular}
  }
  \vspace{3mm}
  \subcaption{Repository code completion}
  \label{tab:repobench_results}
\end{subtable}
\caption{
In (a) the individually optimized attacks are shown to trigger several CWEs when concatenated. In (b) we demonstrate that without accessing the models native tokenizer, significant performance can be achieved using a code-specific tokenizer. In (c) we show that \tool{} preserves code similarity on repository-level completion in RepoBench.
}
\end{figure*}

\subsection{Extended Analysis}
We now investigate how \tool{}'s attacks generalize across CWEs and models and affect repository-level completions.

\paragraph{Multi-CWE Attack}
While \tool{} is mainly developed as a targeted attack, the potential for inducing multiple CWEs simultaneously would exacerbate the posed threat. In \cref{fig:multi_cwe_multi_line}, we investigate the effect of attacking \gptturbo{} with individually optimized attack strings of multiple CWEs together, each included in a new line. For each number of targeted vulnerabilities, we sample 24 unique ordered combinations of CWEs and average the results. This combined attack increases the length of the attack and thus results in a loss of functional correctness, aligning with \cref{fig:num_tokens}. Meanwhile, we observe that the combined attack achieves both a high vulnerability rate and \funcratio{} even at 4 CWEs. Even at 16 simultaneously targeted CWEs, \tool{} achieves an almost $2\times$ higher \vulratio{} than the unattacked engine. These results are both surprising and concerning, as they show that \tool{}'s attacks are composable, without having been explicitly designed for it.

\paragraph{Generalization between Models}
We assess whether attack strings optimized for \scoder{} and \cllama{}, increase the vulnerability of \gptturbo{}. We find that both strings drastically increase $\vulratio$ from $22\%$ to $55\%$ and $59\%$ respectively. Meanwhile, the resulting score is significantly lower than string optimized directly on \gptturbo{} ($73\%$) or for the original models ($80\%$ and $82\%$ on \scoder{} and \cllama{}, respectively). This indicates that the attacks generalize to some degree between models of different sizes and architecture, enabling targeting rate-limited black-box LLMs.

\paragraph{Tokenizer Access}
Under our black-box threat model, the attacker does not have access to the tokenizer of the target engine. Therefore, the attack is optimized in the token space of a proxy tokenizer $\atok$. In our experiments, we use the CodeQwen tokenizer \citep{qwen}, a publicly available tokenizer different from tokenizers of any of the targeted models. In \cref{fig:tokenizer}, we explore the choice of $\atok$, measuring \tool{}'s performance attacking \scoder{} using four different tokenizers: tokenization per Unicode characters, and the GPT-2, CodeQwen, and target model's (StarCoder) tokenizer. We make two key observations. First, the non-code-specific tokenizers (Unicode and GPT-2) lead to low vulnerability rates. Second, the target tokenizer only beats the code-specific proxy $\atok$ in terms of functional correctness on \scoder{}. 
Moreover, as seen in \cref{fig:main}, the proxy tokenizer generalizes to stronger completion engines, incurring virtually no loss even on functional correctness.

\paragraph{Preserving Repository-level completion quality}
In addition to functional correctness in function-level completions in \cref{sec:eval}, we now explore whether \tool{} leads to noticable disturbances on code completion. We focus on repository-level code completion, using RepoBench \citep{liu2023repobench}, a recent benchmark based on GitHub repositories. It measures the similarity of predicted lines to golden completions when given repository-level context. This setting closely aligns with a realistic usage scenario of completion engines. Character-level and syntactic similarity to golden completions are measured using Exact Match (\textsc{EM}), Edit Similarity (\textsc{ES}) and Code Bleu (\textsc{CB}) \citep{ren2020codebleu}.
We randomly choose 333 instances from RepoBench and sample 40 completions of $\engine$ and $\engineadv$. We generate one $\engineadv$ per Python-specific CWE and average the results. We report the ratios of EM, ES and CB in \cref{tab:repobench_results}. The quality of the predicted next line degrades only minimally, if at all, with all scores degrading by at most 16.2\% and on average 8.1\%, overall matching our observations on HumanEval.

\paragraph{Pattern Analysis}
We manually inspect the optimized attack strings to identify patterns. The strings typically contain tokens derived both from initialization strategies and random mutations during optimization. As such, they include a mix of words and code in ASCII and non-ASCII characters, such as non-Latin alphabet letters, symbols from Asian languages, and emojis.
These patterns suggest that, similarly to jailbreak attacks \citep{yong2023low,geiping2024coercing}, our attack partially relies on exploiting low-resource languages and undertrained tokens.
Overall, most attack strings are not easily interpretable by humans. For ethical considerations, we choose not to include the final attack strings publicly in the paper.

\paragraph{More Results and Case Study in Appendix}
In \cref{appsec:additional_experiments}, we provide detailed information on performance per CWE. We further examine sampling temperature choices, as most of our experiments use temperature $0.4$ for optimization and evaluation. In \cref{appsec:cases}, we provide three case studies to illustrate \tool{} attacks.

\section{Discussion}
\label{sec:discussion}
% In this section, we discuss the effectiveness of \tool{}, potential mitigations, limitations, and future work.

\paragraph{\tool{}'s Surprising Effectiveness}
Although our black-box threat model assumes a more restricted attacker than prior attacks \citep{DBLP:conf/ccs/HeV23,DBLP:journals/corr/abs-2301-02344,DBLP:conf/uss/YanWDHLKH24}, \tool{} remains effective both in terms vulnerability rate and functional correctness. This can be attributed to (i) the attack's exploitation of instruction-following capabilities of LLMs, (ii) that many vulnerabilities lie within the learned distribution of LLMs and (iii) that the perturbation introduced by \tool{} is small, allowing capable LLMs to ignore the perturbation usages uncritical to security and generating functionally correct code.

\paragraph{Potential Mitigations}
We appeal to the developers of completion engines to implement mitigations, such as: (i) alerting the user if a string occurs repeatedly at an unusual frequency; (ii) sanitizing prompts before feeding them to the LLM \citep{jain2023baseline}; or (iii) interrupting repeated querying for the purpose of optimizing an attack similar to ours.
%For the latter point, while current code completion engines already implemented rate limits, as evidenced by our success at attacking GitHub Copilot, they are insufficient in preventing \tool{}-style attacks.
We further discuss static analysis, security-inducing comments, and variations of comment scrubbing in \cref{appsec:defenses}.

\paragraph{Limitations and Future Work}
While \tool{} already exposes a concerning vulnerability of today's code completion engines, it incurs some loss on functional correctness of certain completion engines. Stronger attacks could incorporate an explicit optimization objective to preserve functional correctness. Moreover, an interesting future direction would be to extend the attack to other settings such as coding agents \citep{DBLP:conf/iclr/JimenezYWYPPN24} and even more vulnerabilities. 

\section{Related Work}

\paragraph{Code Completion with LLMs}
Transformer-based LLMs excel at solving programming tasks \citep{multiple,zheng2023codegeex}, giving rise to specialized code models such as Codex \citep{DBLP:journals/corr/abs-2107-03374}, CodeGen \citep{DBLP:conf/iclr/NijkampPHTWZSX23}, StarCoder \citep{DBLP:journals/corr/abs-2305-06161} and CodeLlama \citep{DBLP:journals/corr/abs-2308-12950}. LLMs specialized for code completion are trained with a fill-in-the-middle objective \citep{DBLP:journals/corr/abs-2207-14255,DBLP:conf/iclr/FriedAL0WSZYZL23} in order to handle both a code prefix and postfix in their context. Several user studies have confirmed the benefit of LLM-based code completion engines in improving programmer productivity \citep{DBLP:conf/chi/Vaithilingam0G22,DBLP:journals/pacmpl/BarkeJP23}, with such services being used by over a million programmers \citep{developer}.

\paragraph{Security Evaluation of LLM Code Generation}
As code LLMs are increasingly employed, investigating their security implications becomes increasingly imperative. \citet{DBLP:conf/sp/PearceA0DK22} were first to show GitHub Copilot \citep{copilot} frequently generates insecure code. \citet{DBLP:journals/corr/abs-2305-06161,khoury2023secure} extended their evaluation, revealing similar issues in StarCoder and ChatGPT. CodeLMSec \citep{codelmsec} evaluates LLMs' insecure code generation using automatically generated security-critical prompts. However, these works focus on model security only in benign cases, while we examine LLM-based code completion under attack, the worst case from a security perspective.

\paragraph{Random optimization for Jailbreak Attacks}
Random optimization is a common approach to optimize attack strings in jailbreak attacks on LLMs \cite{andriushchenko2025jailbreaking,zou2023universal}. However, the threat model of jailbreak attacks differs significantly from ours. Crucially, in jailbreak attacks, the user is also the attacker and there is no need to build a stealthy attack. As a result, costly random optimization can be applied per prompt, long attack prompts are permissible and there is no need to maintain functional correctness in settings uncritical to security. In contrast, \tool{} derives short, generic attack strings that preserve model performance in uncritical settings.

\paragraph{Attacks on Neural Code Generation}
The common setting in attacks on neural code generation is to assess model robustness by perturbing the entire user input. \citet{cctest-10.1109/ICSE48619.2023.00110,attribution-guided-DBLP:conf/kbse/LiMLXS0L024,DBLP:journals/corr/abs-2312-04730,ren-etal-2024-codeattack} rename variables and functions, among other semantic-preserving perturbations, to trigger functionally incorrect or insecure code completions. While \citet{attribution-guided-DBLP:conf/kbse/LiMLXS0L024,DBLP:journals/corr/abs-2312-04730} target insecure completions, only \citet{DBLP:journals/corr/abs-2312-04730} also ensures preservation of functional correctness. For all methods, their attacks are not suitable for stealthy attacks in our settings, as they assume white-box access or allow expensive search for individual queries. Overall, prior work is designed for users that are interested in assessing LLM robustness, such as model developers. In contrast, we discover a short injection string, the attack comment, that triggers correct but insecure completions over many samples, suitable for our threat model of attacking unassuming users.

Concurrent work by \citet{yang2025tpiatargetspecificpromptinjection} attacks LLMs to trigger insecure completions by injecting code snippets into the RAG context. Their setting differs in three important aspects: First, they leverage white-box model access to optimize their attack. Second, in their RAG setting, larger attack code snippets can be included into the model context. Third, they do not evaluate whether their attack preserves functional correctness and would thus be sufficiently stealthy to succeed under our threat model.

Beyond prompt perturbations, prior attacks achieve increased code vulnerability by interfering either directly with the model weights or its training data \citep{DBLP:conf/uss/SchusterSTS21,DBLP:conf/ccs/HeV23,DBLP:journals/corr/abs-2301-02344,DBLP:conf/uss/YanWDHLKH24}. However, such attacks are unrealistic to be carried out against deployed commercial services.
% In contrast, our attack only requires black-box access to the targeted engine.
% Besides the different threat models, our evaluation covers more CWEs and languages than these works, as discussed in \cref{appsec:additional_experimental_details}.

% In a similar fashion to jailbreaks targeting generic LLMs \citep{zou2023universal,yao2024survey}, DeceptPrompt can synthesize adversarial natural language instructions that prompt LLMs to generate insecure code \citep{DBLP:journals/corr/abs-2312-04730}. 
% However, our work differs from theirs in two significant ways. First, under the threat model of DeceptPrompt, access to the model's full output logits is given, which is often not available for model APIs and commercial engines. \tool{} does not face this limitation and successfully attacks commercial engines, as demonstrated in \cref{sec:eval}. Second, DeceptPrompt only targets a single user prompt at a time. Apart from code generation, \citet{DBLP:conf/icse/YangSH022} leveraged randomized optimization for semantics-preserving transformations to attack code classification models. Both \citet{DBLP:conf/icse/YangSH022} and DeceptPrompt are performed for each input, incurring significant overhead for inference. In contrast, the attack string of \tool{} is derived once and fixed across inputs at inference, thus meeting the real-time requirements of modern code completion. \citep{ren-etal-2024-codeattack,cctest-10.1109/ICSE48619.2023.00110,attribution-guided-DBLP:conf/kbse/LiMLXS0L024,yang2025tpiatargetspecificpromptinjection}
% \todo{fix}
\section{Conclusion}
\label{sec:conclusion}

We presented \tool{}, the first black-box attack capable of manipulating commercial code completion engines to generate insecure code at a high rate while preserving functional correctness. \tool{} inserts a short attack string as comment above the completion line. The string is derived using black-box random optimization that iteratively mutates and selects top-performing attacks. This optimization procedure is further strengthened by a set of diverse initialization strategies. Through extensive evaluation, we demonstrated the surprising effectiveness of \tool{} not only on open-source models but also on real-world production services such as the OpenAI API and GitHub Copilot. Given the broad applicability and high severity of our attack, we advocate for further research into exploring and addressing security vulnerabilities introduced by LLM-based code generation systems.

\section*{Acknowledgements}
This work has been done as part of the EU grant ELSA (European Lighthouse on Secure and Safe AI,
grant agreement no. 101070617) . Views and opinions expressed are however those of the authors
only and do not necessarily reflect those of the European Union or European Commission. Neither
the European Union nor the European Commission can be held responsible for them.

The work has received funding from the Swiss State Secretariat for Education, Research and Innovation (SERI).

\section*{Impact Statement}
\label{sec:ethics_statement}
In this paper, we have introduced \tool{}, the first black-box attack to adversarially steer (commercial) code completion engines towards generating insecure code. As our attack can be potentially developed even by an attacker with notably low resources, and deployed on commercial services exploiting well-known vulnerabilities of, for instance, IDE plug-in marketplaces; we have made careful steps to ensure that our research process and publication of our results is aligned with the ethical responsibilities carried by the potential harms of \tool{}. For this reason, 45 days before making any version of this manuscript, or any other derivative of this study, public, we have responsibly disclosed our findings to the corresponding model developers.
Due to ethical concern, we did not include any concrete optimized attack strings in this paper, nor in any supplementary material. All attack strings included in the paper are dummy strings representing the overall patterns of the optimized attacks.
Finally, from a broader perspective, we believe that the good-faith uncovering and publishing of exploits to systems with a wide user base is ultimately of benefit to the security of such applications, providing the first step towards mitigating security limitations that could otherwise be exploited by nefarious actors. 
We release our dataset and code implementation to allow the research community to replicate and advance our findings.

\bibliography{paper}
\bibliographystyle{icml2025}

\clearpage
\appendix
\onecolumn
\section*{Appendix}
\section{Extended Details on Experimental Setup}
\label{appsec:additional_experimental_details}

We now give additional details about our implementation, hyperparameters, and vulnerability dataset.

\paragraph{Implementation and Hyperparameters}
The results in our main experiments (\ie{}, \cref{fig:main}) are obtained with the following configuration: attack comment positioned in the line above the completion point, optimization and initialization combined, CodeQwen tokenizer \citep{qwen}, pool size $\poolSize=20$, and, following \citet{DBLP:conf/ccs/HeV23}, sampling temperature during optimization and evaluation $0.4$.
The number of tokens in the attack string is set to $\attackLen=5$ for all engines and vulnerabilities except: $\attackLen=10$ for Copilot on five vulnerabilities, and $\attackLen=15$ for Copilot on one vulnerability. We select these hyperparameters according to our experiments on the validation datasets $\funcval$ and $\valdata$ and the ablations presented in \cref{sec:eval} and \cref{appsec:additional_experiments}. During optimization, for each candidate string, we sample $16$ completions per task to approximate $\vulratio$ in \cref{eq:vul}. As running CodeQL during optimization would be prohibitively slow, we use approximate rule-based classifiers to determine if a completion is vulnerable. As the final scores are computed using accurate assessment via CodeQL, this confirms that such classifiers are accurate enough on our training samples. As we mention in \cref{appsec:defenses}, such manually written classifiers would likely be a tool of preferred choice for attackers trying to introduce novel vulnerabilities.
Finally, when mutating attack strings we forbid a set of problematic tokens: those including new lines and special tokens, such as \texttt{<|endoftext|>}.
As models, we used \texttt{gpt-3.5-turbo-instruct-0914} for \gptturbo{} and the standard GitHub Copilot plugin as of June 2024.

\paragraph{Vulnerability Dataset}
Our vulnerability dataset consists of 16 CWEs across 5 programming languages. We show an overview of these vulnerabilities, their MITRE vulnerability rank, and a short description in \cref{tab:cwe_overview}.
Further, for each CWE, we construct $12$ realistic completion tasks using three different sources: (i) we incorporate all suitable tasks from the dataset of \citet{DBLP:conf/sp/PearceA0DK22}, (ii) we search GitHub for code that contains or fixes each specific CWE to collect real-world samples, and (iii) when the above sources do not yield sufficient samples, we leverage GPT-4 to generate additional samples based on detailed descriptions of the CWEs. We invested significant effort in reviewing and revising the samples to ensure high quality. Our primary objective during this process was to ensure diversity, realism, and sufficient context for the completion engines to generate functional code.

\begin{wraptable}[5]{r}{0.44\textwidth}
  \vspace{-7mm}
  \begin{minipage}{0.44\textwidth}
    \footnotesize
    \centering
    \begin{tabular}{@{}lrr@{}}
      \toprule
      & \#CWEs & \#LANGs \\
      \midrule
      \citet{DBLP:conf/uss/SchusterSTS21} & 3 & 1 \\
      \citet{DBLP:conf/sp/PearceA0DK22} & 18 & 2 \\
      \citet{DBLP:conf/ccs/HeV23} & 9 & 2 \\
      \citet{DBLP:journals/corr/abs-2301-02344} & 4 & 1 \\
      \citet{DBLP:conf/uss/YanWDHLKH24} & 3 & 1 \\
      Our Work & 16 & 5 \\
      \bottomrule
    \end{tabular}
  \end{minipage}
\end{wraptable}
In the table on the right, we compare the evaluation scope of our work with prior studies. Our work covers a broader or comparable range of CWEs and programming languages, highlighting the thoroughness of our evaluation. This underscores the potential of our dataset as a valuable contribution to the community.
\begin{table}[!b]
  \centering
  \small
  \def\arraystretch{1.1}
  \setlength\tabcolsep{10pt}
  \caption{Overview of the CWEs studied in this paper and the size of the corresponding dataset.}
  \vspace{-2mm}
  \label{tab:cwe_overview}
  \begin{tabular}{@{}p{1mm}lcccr@{}}
    \toprule
    \# & CWE & Language & Top-25 CWE Rank & Avg LoC & Max LoC  \\
    \midrule
    20 & Improper Input Validation & Python & \#6 &  16 &  22 \\
    22 & Path Traversal & Python & \#8 & 14 &  28  \\
    77 & Command Injection & Ruby & \#16 & 9 &  19  \\
    78 & OS Command Injection & Python & \#5 & 15 &  30  \\
    79 & Cross-site Scripting & JavaScript & \#2 & 19 &  27  \\
    89 & SQL Injection & Python & \#3 & 19 &  32  \\
    90 & LDAP Injection  & Python & -- & 23 &  33  \\
    131 & Miscalculation of Buffer Size & C/C++ & -- & 22 &  35  \\
    193 & Off-by-one Error & C/C++ & -- & 26 &  54  \\
    326 & Weak Encryption & Go & -- & 34 &  75  \\
    327 & Faulty Cryptographic Algorithm & Python & -- & 14 &  34  \\
    416 & Use After Free & C/C++ & \#4 & 18 &  22  \\
    476 & NULL Pointer Dereference & C/C++ & \#12 & 22 &  68  \\
    502 & Deserialization of Untrusted Data & JavaScript & \#15 & 14 &  18  \\
    787 & Out-of-bounds Write & C/C++ & \#1 & 21 &  52  \\
    943 & Data Query Injection & Python & -- & 25 &  31  \\
    \bottomrule
  \end{tabular}
\end{table}
\newpage

\diff{\paragraph{CodeQL as Vulnerability Judgment}
Since our evaluation of vulnerabilities relies on CodeQL as a judgment function, we need to ensure that its judgment is trustworthy in our setting. To reduce false positives, we select only relevant CodeQL queries for each CWE. We further manually evaluate the precision of CodeQL on $\testdata$, by sampling 50 instances from diverse settings, covering all models, CWEs, and the presence of none, Init-only, and optimized attack strings. We find that CodeQL exhibits high precision on our dataset, with $98\%$ actual vulnerabilities reported.}

\begin{wraptable}[15]{r}{.48\textwidth}
  \centering
  \vspace{4mm}
  \setlength{\tabcolsep}{4pt}
  \renewcommand{\arraystretch}{0.9}
  \caption{Comparison of \funcratio{} and \funcratioten{} between manually translated and GPT-4-generated reference solutions.}
  \vspace{-2mm}
  \label{tab:human_eval_translation}
  \begin{tabular}{lccccc}
    \toprule
    & \multicolumn{2}{c}{\funcratioone{}} & \multicolumn{2}{c}{\funcratioten{}} \\
    \cmidrule(lr){2-3} \cmidrule(lr){4-5}
    Model & Manual & GPT-4 & Manual & GPT-4 \\
    \midrule
    \scoder & 78.0 & 74.9 & 99.8 & 97.9 \\
    \cllama & 88.2 & 87.7 & 99.8 & 99.7 \\
    \scodertwoThreeB & 89.5 & 90.3 & 100.2 & 99.5 \\
    \scodertwoSevenB & 87.0 & 85.2 & 99.8 & 99.3 \\
    \scodertwoFifteenB & 94.6 & 96.1 & 100.5 & 100.1 \\
    \bottomrule
  \end{tabular}
\end{wraptable}
\paragraph{Validation of GTP-4-Generated HumanEval Solutions}
For our evaluation of functional correctness, we evaluated the effect of \tool{} on infilling tasks generated from GPT-4 generated solutions to HumanEval in other languages than Python.
While translations of HumanEval prompts exist, \eg{} in \citep{multiple}, only the dataset HumanEval-X \citep{zheng2023codegeex} contains human-written translations of the reference solutions for some languages, and we found no manual translation of Ruby.
As we preferred to treat the different languages equally, we decided to generate solutions for all non-Python languages using GPT-4.
To validate our results, we compare our results to manually translated samples in languages C++ and JavaScript.
The comparison of the obtained \funcratio{} is displayed in \cref{tab:human_eval_translation}, confirming that the obtained results are similar between manually translated and GPT4-generated reference solutions.

\section{Details on Attack Optimization}
\label{appsec:attack_pseudocode}

In this section we provide more detailed pseudocode and descriptions of the attack optimization conducted by \tool{}, specifically, we provide the implementations of the \pickfunc{} and \mutatefunc{} functions.

\paragraph{Selection}
The function \pickfunc{} is used to select the $n$ top-performing attack strings from a given pool.
We present its details in \cref{algo:pick}.
For each attack string $\attackParam \in \attackPool$ (Line 3), we first construct a malicious completion engine $\engineadv$ with $\attackParam$ (Line 4).
Then, at Line 5, sampling completions to the tasks in $\vuldata$, we estimate the $\vulratio(\engineadv)$ when attacked using the current $\attackParam$.
Finally, in Line 7, we pick and return the $n$ best attack strings according to the vulnerability scores collected in $\mathcal{V}$. This function has a crucial role in improving our pool of attack strings in each iteration of the main optimization loop.

\paragraph{Mutation}
The function \mutatefunc{} is used in the main optimization loop of \cref{algo:toplevel} to randomly alter the attack strings in the candidate pool. It is an important step for \tool{}'s optimization algorithm to discover stronger attack strings.
We present the internals of \mutatefunc{} in \cref{algo:mutate}.
First, using the attacker's tokenizer $\atok$, we tokenize $\attackParam$ (Line 2).
Note that to comply with our black-box threat model, we assume that the attacker obtains $\atok$ independently, thus it does not necessarily match the tokenizer of the targeted engine $\engine$.
Next, in Line 3, we uniformly sample the number of tokens $k$ that will be mutated in $\attackParam$.
Then, in Line 4, we randomly sample $k$ positions $\mathcal{I}$ to mutate.
In Lines 5 and 6, for each position index $i \in \mathcal{I}$, we mutate $\mathbf{t}[i]$ by replacing it with a token sampled uniformly at random from the vocabulary of $\atok$.
Finally, we return the detokenized mutated string.

\begin{figure*}
    \hfill
    \begin{minipage}{.45\textwidth}
        \input{algorithms/pick.tex}
    \end{minipage}
    \hfill
    \begin{minipage}{.45\textwidth}
        \input{algorithms/mutate.tex}
    \end{minipage}
    \hfill
\end{figure*}

\section{Initialization Strategy Details}
\label{appsec:initialization_scheme_details}

In this section, we give extended details on each initialization scheme used in \tool{}. A high-level description of their invocation has been introduced in \cref{sec:attack-init}.

\paragraph{Random Initialization}
We increase the diversity of our initialization by generating random attack strings. We achieve this by randomly sampling tokens from the attacker's tokenizer $\atok$ and concatenating them into strings. Note that such generated strings are not usually completely random characters, but feature some structure based on the size and content of the tokenizer dictionary. An example for such a string $\attackParam$ is ``\texttt{éd senior \raisebox{-0.5mm}{\includegraphics[width=0.5em]{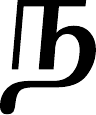}} sp cuts}", which includes complete words and Unicode characters and was generated by sampling tokens at random from the CodeQwen tokenizer \citep{qwen}.

\paragraph{TODO Initialization}
We initialize the attack string $\attackParam$ to ``\texttt{TODO: fix vul}'' to indicate that the code to be completed was marked, \eg{}, by a human developer, to contain a security vulnerability. If the completion engine is aware of potential vulnerabilities or has picked up similar code snippets containing review notes and insecure code, we expect it to be steered toward generating the corresponding insecure code.

\paragraph{Security-critical Token Initialization}
We observe that, for a wide range of vulnerabilities, there exist critical tokens that decide the security of the whole program. For instance, consider the following implementation of a database query using securely parameterized SQL:
\texttt{cursor.execute(\textquotesingle SELECT ... WHERE id=}\textcolor{mygreen}{\texttt{?\textquotesingle,}}\texttt{ q\_id)}.
Here, \texttt{q\_id} is an untrusted user input and the \textcolor{mygreen}{\texttt{?\textquotesingle,}} parametrization makes sure that any potentially dangerous characters in \texttt{q\_id} are escaped. In contrast, an insecure implementation would be:
\texttt{cursor.execute(\textquotesingle SELECT ... WHERE id=}\textcolor{myred}{\texttt{\textquotesingle\ +}}\texttt{ q\_id)},
where the untrusted input is directly concatenated to the query without any checks. As such, the security-critical tokens are ``\textcolor{mygreen}{\texttt{?\textquotesingle,}}'' and ``\textcolor{myred}{\texttt{\textquotesingle\ +}}''. The concrete tokens for each CWE can be extracted directly using the training dataset and secure and insecure completions by computing the textual difference. We exploit this pattern to create an initialization strategy yielding strings of the format ``\texttt{use \{insecure tokens\}}'' and ``\texttt{don\textquotesingle t use \{secure tokens\}}''. For the above example of SQL injection in Python, we would create initial attack strings ``\texttt{use }\textcolor{myred}{\texttt{\textquotesingle\ +}}'' and ``\texttt{don\textquotesingle t use }\textcolor{mygreen}{\texttt{?\textquotesingle,}}''.

\paragraph{Inversion Initialization}
\begin{wrapfigure}[13]{r}{0.48\textwidth}
  %\vspace{-4mm}
  \begin{minipage}{0.48\textwidth}
    \input{figures/inversion_example/figure.tex}
  \end{minipage}
\end{wrapfigure}
\tool{} works by inserting a comment such that insecure code gets generated by the underlying completion engine. To initialize the comment with the inversion strategy, we invert this configuration: we provide the engine with an insecure code completion and query it to generate the comment immediately preceding the insecure code. A concrete example is provided in \cref{fig:inversion-example}. Here the model is expected to complete the part marked by ``\texttt{\{attack string $\sigma$\}}'' and is provided with an insecure usage of the \texttt{md5} function as opposed to the model default of the secure \texttt{sha256} function. This strategy exploits the engine's learned relationship between vulnerable code and related comments in the distribution of its training data.

\paragraph{Sanitizer Initialization}
Many injection-style vulnerabilities, such as cross-site scripting, can be mitigated by applying specific sanitization functions on potentially unsafe objects. For example, the \texttt{escape} function from the \texttt{escape-html} library \citep{escape} can be used to safely encode user inputs that could be interpreted as valid HTML tags before they are displayed on web pages (cf. CWE-79). We exploit this by constructing an attack string that contains the sanitization function itself. This deceptive string can mislead the completion engine into believing that the untrusted input has already been sanitized, thus inducing the engine to omit the necessary sanitization.

Given that the attacker may not know in advance which variable name should be sanitized, we design the attack string to be generic, targeting a variable \texttt{x}. As a result, the attack string is formulated as ``\texttt{x = \{sanitizer\}(x)}'', where \texttt{\{sanitizer\}} is replaced by the actual sanitization function, such as \texttt{escape}. Concretely, the sanitizer initialization string $\attackParam$ in the JavaScript CWE-79 setting of our experiments is ``\texttt{x = escape(x)}".

\section{Additional Experiments}
\label{appsec:additional_experiments}

In this section, we present additional details on \tool{}'s impact per CWE and temperature ablations.

\paragraph{Attack Performance per CWE}
In \cref{fig:breakdown_gpt35}, we show our main results on \cllama{} broken down per CWE. We order the CWE by the final vulnerability score of \tool{}.
First of all, we observe that our attack manages to increase the vulnerability rate of the generated programs across all vulnerabilities, except for CWE-079-js and CWE-020-py where the original completion engine already has a high vulnerability rate.
In particular, our attack manages to trigger a vulnerability rate of over $90\%$ on more than a third of all examined CWEs. Remarkably, in several cases \tool{} manages to trigger such high attack success rates even though the base model had a vulnerability rate of close to zero.
Further, we observe that while the $\funcratioone$ of \cllama{} averaged across all 16 vulnerabilities is $89\%$ (see \cref{fig:main}), this average is composed of a bimodal distribution. Attacks targeting certain vulnerabilities have larger relative impact on functional correctness ($\geq 25\%$), while others have almost no impact.

\begin{figure*}
  \centering
  \resizebox*{\columnwidth}{!}{
  \begin{tikzpicture}
    \centering
    \begin{groupplot}[
      height=3.6cm, width=18cm,
      /pgf/bar width=0.32cm,
      axis x line*=bottom, axis y line*=left, enlarge x limits=true,
      xtick={0, 1},
      xticklabel style={yshift=-0.8mm, font=\footnotesize, align=center},
      ybar=3.5pt,
      ymin=0, ymax=110, ytick={0, 25, 50, 75, 100}, yticklabels={0, 25, 50, 75, 100},
      ymajorgrids, major grid style={draw=black!20}, tick align=inside,
      yticklabel style={font=\scriptsize}, tickwidth=0pt,
      y axis line style={opacity=0},
      nodes near coords={\pgfmathprintnumber[precision=0]{\pgfplotspointmeta}},
      every node near coord/.append style={font=\scriptsize},
      group style={group size=1 by 2, horizontal sep=30pt, vertical sep=30pt},
      legend style={
        at={(0.5,1.1)},  % Position of the legend
        anchor=south,  % Anchor point of the legend
        legend columns=-1,  % Number of columns in the legend (-1 means that all entries are in one row)
        draw=none,
        column sep=0.1cm,
        /tikz/column 2/.style={
          column sep=0.7cm,
        },
        /tikz/column 4/.style={
          column sep=0.7cm,
        },
      },
      legend image code/.code={%
        \draw[#1, draw=mydrawgray] (0cm,-0.125cm) rectangle (0.5cm,0.125cm);
      },
      legend cell align={left},
    ]

    % ['cwe-131_cpp', 'cwe-943_py', 'cwe-787_cpp', 'cwe-327_py', 'cwe-502_js', 'cwe-089_py', 'cwe-416_cpp_p', 'cwe-476_cpp', 'cwe-022_py', 'cwe-090_py', 'cwe-078_py', 'cwe-077_rb', 'cwe-193_cpp', 'cwe-079_js', 'cwe-326_go', 'cwe-020_py']

    \nextgroupplot[
      xmin=0.2, xmax=6.8,
      xtick={0, 1, 2, 3, 4, 5, 6, 7},
      xticklabels={CWE 131, CWE 943, CWE 787, CWE 327, CWE 502,  CWE 089, CWE 416, CWE 476},
    ]
      \addplot [draw=mydrawgray, line width=0.7pt, fill=baselineVulColor] table [x=vul, y=baselineVR, col sep=comma] {figures/vul_breakdown/data_row1.csv};
      \addlegendentry{\small $\vulratio(\engine)$}

      \addplot [draw=mydrawgray, line width=0.7pt, fill=optVulColor] table [x=vul, y=optVR, col sep=comma] {figures/vul_breakdown/data_row1.csv};
      \addlegendentry{\small $\vulratio(\engineadv)$}

      \addplot [draw=mydrawgray, line width=0.7pt, fill=pass1Color, postaction={pattern=north east lines, pattern color=white}] table [x=vul, y=pass@1, col sep=comma] {figures/vul_breakdown/data_row1.csv};
      \addlegendentry{\small $\funcratioone(\engineadv, \engine)$}

      % \addplot [draw=mydrawgray, line width=0.7pt, fill=pass10Color] table [x=vul, y=pass@10, col sep=comma] {figures/vul_breakdown/data_row1.csv};    

    \nextgroupplot[
      xmin=0.2, xmax=6.8,
      xtick={0, 1, 2, 3, 4, 5, 6, 7},
      xticklabels={CWE 022, CWE 090, CWE 078, CWE 077, CWE 193, CWE 079, CWE 326, CWE 020},
    ]
      \addplot [draw=mydrawgray, line width=0.7pt, fill=baselineVulColor] table [x=vul, y=baselineVR, col sep=comma] {figures/vul_breakdown/data_row2.csv};

      \addplot [draw=mydrawgray, line width=0.7pt, fill=optVulColor] table [x=vul, y=optVR, col sep=comma] {figures/vul_breakdown/data_row2.csv};

      \addplot [draw=mydrawgray, line width=0.7pt, fill=pass1Color, postaction={pattern=north east lines, pattern color=white}] table [x=vul, y=pass@1, col sep=comma] {figures/vul_breakdown/data_row2.csv};

      % \addplot [draw=mydrawgray, line width=0.7pt, fill=pass10Color] table [x=vul, y=pass@10, col sep=comma] {figures/vul_breakdown/data_row2.csv};

    \end{groupplot}
  \end{tikzpicture}}
  \vspace{-5mm}
  \caption{Breakdown of our \tool{} attack applied on \cllama{} over different vulnerabilities.}
  \label{fig:breakdown_gpt35}
\end{figure*}

\begin{figure}[t]
  \hfill
  \centering
  \begin{subfigure}[b]{0.36\textwidth}
        \centering
  \resizebox*{\columnwidth}{!}{
  \begin{tikzpicture}
    \centering
    \begin{axis}[
      height=5cm, width=7cm,
      axis lines=middle, axis x line*=bottom, axis y line*=left, enlarge x limits=false,
      tick align=inside, minor tick style={draw=none},
      xticklabel style={font=\small}, yticklabel style={font=\small},
      ymin=50, ymax=92, ytick={55, 63, 71, 79, 87}, 
      xmin=-0.1, xmax=1.1, xtick={0, 0.2, 0.4, 0.6, 0.8, 1.0}, xticklabels={0, 0.2, \textbf{0.4}, 0.6, 0.8, 1.0},
      line width=1pt,
      x label style={at={(axis description cs:0.5,-0.13)}, anchor=north},
      xlabel={\small Optimization temperature},
      legend style={
        at={(0.5,1.03)},  % Position of the legend
        anchor=south,  % Anchor point of the legend
        legend columns=-1,  % Number of columns in the legend (-1 means that all entries are in one row)
        draw=none,
        column sep=0.1cm,
        /tikz/column 2/.style={
          column sep=0.4cm,
        },
      },
      legend cell align={left},
    ]
      \addplot [mark=*, mark size=2.5pt, draw=optVulColor, mark options={fill=optVulColor}] table [x=trainTemp, y=vulRatio, col sep=comma] {figures/train_temp/data.csv};
      \addlegendentry{\small $\vulratio(\engineadv)$}

      \addplot [mark=triangle*, mark size=3.1pt, draw=pass1Color, mark options={fill=pass1Color}] table [x=trainTemp, y=pass@1, col sep=comma] {figures/train_temp/data.csv};
      \addlegendentry{\small $\funcratioone$}
    \end{axis}
  \end{tikzpicture}}
  \vspace{-4mm}
  \subcaption{Varying optimization temperatures with fixed evaluation temperature $0.4$.}
  \label{fig:train_temp}
  \end{subfigure}
  \hfill
  \begin{subfigure}[b]{0.40\textwidth}
        \centering
  \resizebox*{\columnwidth}{!}{
  \begin{tikzpicture}
    \centering
    \begin{axis}[
      height=6cm, width=9cm,
      axis lines=middle, axis x line*=bottom, axis y line*=left, enlarge x limits=false,
      tick align=inside, minor tick style={draw=none},
      xticklabel style={font=\large}, yticklabel style={font=\large},
      ymin=50, ymax=100, ytick={55, 65, 75, 85, 95}, 
      xmin=-0.1, xmax=1.1, xtick={0, 0.2, 0.4, 0.6, 0.8, 1.0}, xticklabels={0, 0.2, \textbf{0.4}, 0.6, 0.8, 1.0},
      line width=1.3pt,
      x label style={at={(axis description cs:0.5,-0.13)}, anchor=north},
      xlabel={\large Evaluation temperature},
      legend style={
        at={(0.46,1.05)},  % Position of the legend
        anchor=south,  % Anchor point of the legend
        legend columns=-1,  % Number of columns in the legend (-1 means that all entries are in one row)
        draw=none,
        column sep=0.1cm,
        /tikz/column 2/.style={
          column sep=0.4cm,
        },
      },
      legend cell align={left},
    ]
      \addplot [mark=*, mark size=3.1pt, line width=1.5pt, draw=optVulColor, mark options={fill=optVulColor}] table [x=evalTemp, y=vulRatio, col sep=comma] {figures/eval_temp/data.csv};
      \addlegendentry{\small $\vulratio(\engineadv)$}

      \addplot [mark=triangle*, mark size=3.5pt, line width=1.5pt, draw=pass1Color, mark options={fill=pass1Color}] table [x=evalTemp, y=pass@1, col sep=comma] {figures/eval_temp/data.csv};
      \addlegendentry{\small $\funcratioone$}

      \addplot [mark=diamond*, mark size=3.5pt, line width=1.5pt, draw=pass10Color, mark options={fill=pass10Color}] table [x=evalTemp, y=pass@10, col sep=comma] {figures/eval_temp/data.csv};
      \addlegendentry{\small $\funcratioten$}
    \end{axis}
  \end{tikzpicture}}
  \vspace{-1mm}
  \subcaption{Varying evaluation temperatures with a fixed attack.}
  \label{fig:eval_temp}
  \end{subfigure}
  \hfill
  \vspace{-0mm}
\caption{
  In (a) we evaluate the ideal temperature range for evaluating attack strings during optimization to be [0.2 - 0.4]. In (b), it can be seen that the attack is most effective on targeted engines with low temperatures.
}
\end{figure}

\paragraph{Optimization Temperature}
Recall that, at Line 5 of \cref{algo:pick}, we evaluate the vulnerability rate of a malicious completion engine, either on the training set $\traindata$ or the validation set $\valdata$. This assessment requires sampling from the targeted engine, for which temperature plays a critical role in controlling the sample diversity. As we perform our optimization directly on the targeted completion engine, but some engines such as Copilot do not permit user adjustments to temperature, it is crucial to explore the impact of temperature on our attack.
In \cref{fig:train_temp}, we explore temperatures ranging from $0$ to $1.0$ during optimization. Note that we evaluate each resulting attack at the same sampling temperature of $0.4$ for fair comparison.
First, we observe that our attack achieves a non-trivial vulnerability rate at any optimization temperature, which implies that even APIs where this parameter cannot be set are vulnerable to \tool{}.

Next, we can see that there is an ideal range of temperature values ($0.2-0.4$) for the model on which the optimization is conducted where the attack is highly successful, \ie, it achieves high vulnerability rate while retaining a good amount of functionality in the completions. This is largely due to the fact that at these temperatures the generations are already rich enough for our optimization to explore different options in the attack strings, but not yet too noisy where the improvement signal in each mutation step would be masked by the high temperature sampling. Based on this insight, we pick a temperature of $0.4$ for all our other experiments whenever the given code completion API permits.

\paragraph{Evaluation Temperature}
Additionally to the temperature during optimization, of equal importance is to consider the temperature under which the attack is deployed, \ie, the temperature during evaluation. Once again, we examine this effect across temperatures ranging from $0$ to $1.0$ in \cref{fig:eval_temp}. We can observe that at low temperatures, typically preferred for code generation (\eg, $0.0-0.4$), \tool{} achieves a high vulnerability rate and functional correctness. As temperature increases, the vulnerability rate of the attack decreases, as also observed by \citet{DBLP:conf/ccs/HeV23}. However, the vulnerability rate still remains high, indicating that the attack continues to pose a serious threat. In terms of functional correctness, $\funcratioten$ is a more relevant metric for high temperature \citep{DBLP:journals/corr/abs-2107-03374} and the attack can maintain $\funcratioten$ across different temperatures.
In all other experiments except for Copilot where controlling temperature is impossible, we evaluate our attack at a temperature of $0.4$, which is a middle point and also aligns with the setup of \citet{DBLP:conf/ccs/HeV23}.

\section{Case Studies}
\label{appsec:cases}

In \cref{fig:opt_only}, we show that both initialization and optimization alone achieve good attack performance and that they are even more effective when combined. Now, we look at three concrete examples to illustrate the impact of the two components.

% \begin{wrapfigure}{r}{0.48\textwidth}
%   % \vspace{-1.0cm}
%   \begin{minipage}{0.48\textwidth}
%     \input{figures/case_studies/figure1.tex}
%   \end{minipage}
% \end{wrapfigure}
For certain cases, the initialization is already very effective and does not need further optimization. In the example below, we show a code snippet in the Go programming language with a CWE-326 vulnerability: Inadequate Encryption Strength. Due to increased computer capabilities, 1024-bit RSA keys are considered insecure. In this case, our security-critical token initialization creates an attack string ``\texttt{use 1024}''. \cllama{} follows the instruction in the attack string and already produces insecure code with a $100\%$ probability. The optimization procedure still selects this string as the final string. Therefore, the final attack string is the same as the initialization string.
\begin{minipage}{\textwidth}
\begin{lstlisting}[language=Go]
privateKey, err := ParsePemKey(a.PrivateKey)
b, _ := pem.Decode([]byte(a.RootCertificate))
cert, err := x509.ParseCertificate(block.Bytes)
// (*@\mytextcolor{attackString}{attack string $\attackParam$}@*)
key, err := rsa.GenerateKey(rand.Reader, 1024)
if err != nil {
  err = errors.New("Failed to generate rsa key")
  return
}
\end{lstlisting}
\end{minipage}

% \begin{wrapfigure}{r}{0.48\textwidth}
%   % \vspace{-1.0cm}
%   \begin{minipage}{0.48\textwidth}
%     \input{figures/case_studies/figure2.tex}
%   \end{minipage}
% \end{wrapfigure}
Our optimization can also refine initial strings to produce significantly stronger attacks. Below is a Python example of CWE-022: Path Traversal. The \texttt{send\_file} function is vulnerable as an attacker can use it to access files outside of the intended directory. Instead, the secure version should use the \texttt{send\_from\_directory} function. In this case, our initialization creates an attack string ``\texttt{use send\_file}'' with a good vulnerability rate of $65\%$ for \gptturbo{}. The optimization refines the attack string by keeping \texttt{send\_file} while replacing other tokens with non-ASCII tokens. As a result, the attack strength is increased to $96\%$. The new tokens in the optimized attack string do not have a human-interpretable meaning and thus it would be difficult to obtain these tokens without our optimization procedure.
\begin{minipage}{\textwidth}
\begin{lstlisting}[language=Python]
@web_app.route('/js/<f:f>')
def send_js_files(f):
  js_dir = "frontend/js"
  print(f[:-4])
  # (*@\mytextcolor{attackString}{attack string $\attackParam$}@*)
  return flask.send_file(os.path.join(js_dir, f))

@web_app.route("/")
def main_page():
  return flask.render_template("index.html")
\end{lstlisting}
\end{minipage}

% \begin{wrapfigure}{r}{0.48\textwidth}
%   % \vspace{-1.0cm}
%   \begin{minipage}{0.48\textwidth}
%     \input{figures/case_studies/figure3.tex}
%   \end{minipage}
% \end{wrapfigure}
Finally, we demonstrate how our optimization substantially improves a randomly initialized attack string. The C/C++ example below contains a CWE-131 vulnerability: Incorrect Calculation of Buffer Size. The \texttt{malloc} function is used to allocate memory for the new string, but the size is too small. This leaves the code susceptible to a buffer overflow attack, which could lead to anything from program crashes to malicious code execution. Initialization creates only attacks with a low vulnerability rate of $\leq 14\%$ for \scoder{}. The final attack string originates from a randomly initialized string with non-ASCII tokens and brings the vulnerability rate up to $63\%$. Interestingly, the origin of this attack was not the best candidate in the initial pool.

\begin{minipage}{\textwidth}
\begin{lstlisting}[language=C]
char* append_exclamation(const char *input) {
  char *new_input;
  // (*@\mytextcolor{attackString}{attack string $\attackParam$}@*)
  new_input = (char *) malloc(strlen(input));
  if (new_input == NULL) {
    fprintf(stderr, "Failed to allocate memory");
    exit(1);
  }
  strcpy(new_input, input);
  new_input[strlen(input)] = '!';
  return new_input;
}
\end{lstlisting}
\end{minipage}

% \section{Naming}
% 
% The name for \tool{} is an abbreviation of \textbf{In}jecting \textbf{S}ecurity \textbf{E}vading \textbf{C}omments, a reference to the fact that the method relies on the injection of short comments to decrease security (increase vulnerability) of generated code in LLMs.
\section{Discussion of Defenses}
\label{appsec:defenses}

In this section we discuss possible defenses against \tool{}, such as adding comments to counter the effect of \tool{}, scrubbing all comments from prompts, and deploying static analysis in production.

\paragraph{Static Analysis}
While static analysis~\cite{codeql} is suitable for the purpose of our evaluation, it is not implied that it could reliably detect and prevent vulnerabilities generated by LLMs.
This is because, as also discussed in \cref{sec:eval}, our evaluation handles known vulnerabilities, allowing us to utilize specialized CodeQL queries tailored for each individual scenario and thereby achieving high accuracy.
In contrast, effective vulnerability detection requires a more general approach capable of addressing various types of undiscovered vulnerabilities.
First, \tool{} can be extended to trigger vulnerabilities that are not covered or difficult to detect for static analysis.
\citet{DBLP:conf/uss/YanWDHLKH24} has demonstrated the feasibility of such an evasive attack in a white-box setting.
It is an interesting topic for future research to adapt it to the black-box setting of \tool{}.
Second, even for known and detectable CWEs, static analysis tools are rarely configured appropriately \citep{charoenwet2024empirical}, suffer from poor explanations for discovered vulnerabilities \citep{nachtigall2019explaining}, and lack actionable advice for mitigation \citep{nachtigall2023evaluation}.
This results in static analysis being much less prevalent in practice than might be expected \citep{ryan2023unhelpful}, with Copilot-generated vulnerable code already being found in public GitHub repositories \citep{DBLP:journals/corr/abs-2310-02059}.

\paragraph{Security Comments}
We investigate whether adding additional comments can mitigate our attack when such comments instruct the model to generate secure code. We insert \texttt{This code should be secure} in the line above the INSEC attack string, using the attack string optimized without the presence of the comment. This setting is the worst case from an attacker's perspective since they could not adapt to the deployed defense. On \gptturbo{}, averaged over all CWEs, this slightly decreases the vulnerability ratio from $76\%$ to $62\%$. This score still largely exceeds the baseline ratio of only $22\%$. 
This result is not surprising, as previous work has found that usual, unoptimized comments are insufficient to steer models towards secure code generation \citep{pmlr-v235-he24k,liu2024solitary}.
As noted, beyond this, the attacker may adapt to such a deployed defense by re-running the attack string optimization, taking into account the presence of such a security-inducing comment.
Exploration of the interaction between opposing optimization schemes for and against code security would pose an interesting topic of future research.

\paragraph{Comment Scrubbing}
In contrast, we investigate the scrubbing of all comments from code as a possible avenue for defense. We note that code models rely on comments to steer their generations \citep{anonymous2024steering,song2024codeneedscommentsenhancing} and suspect that removal of comments generally reduces performance on standard tasks. We evaluate this experimentally by removing all comments from the HumanEval dataset and replacing them with stub comments, before requesting fill-in completion, for \scoder{}, the \scodertwoFamily{}, and \gptturbo{}. We observe an overall \funcratioone of only $89.6\%$ compared to vanilla completions, matching the decrease in functionality due to \tool{}. As developers are usually not willing to sacrifice functional correctness for security \citep{pmlr-v235-he24k}, and may get frustrated at the lack of steerability of the LLM, we suspect that straightforward removal is not a suitable defense.

\paragraph{Removing non-ASCII characters}
We run our attack optimization excluding non-ASCII characters on \gptturbo{}, only sampling random initializations and mutations from ASCII-only tokens. We observe that attacks under such a constrained setting are still successful, achieving an increase of vulnerability rate from $17.1\%$ to $73.1\%$, similar to the $72.5\%$ in the unconstrained setting. Meanwhile, functional correctness is preserved with \passone{} (\passten) of $98.3\%$ ($99.9\%$). We conclude that ASCII-only removal is generally not an effective defense against \tool{}.

\paragraph{Perplexity Based Defenses}
Recent work suggested perplexity based defenses to prevent jailbreak attacks \citep{jain2023baseline}, rejecting providing completions to queries with high perplexity.
Since our attack is indiscriminately inserted into all user queries, perplexity filters designed to reject security-relevant, attacked queries might also reject benign queries for functional code completion, undermining the code completion engine's utility. The necessity to maintain functional correctness is a key difference between our setting and jailbreak defenses.

To demonstrate this experimentally, we examine perplexity filters as employed by \citet{jain2023baseline}. First, we choose a rejection threshold that maximizes the F1 score of detecting attacked prompts in the training and validation set of our vulnerability dataset, achieving recall of over $89\%$ on the test set. Applying this filter on the functional correctness dataset drastically decreases correctness for benign prompts, with \funcratioone{} (\funcratioten) of less than $29.8\%$ ($29.4\%$), rendering the defense impractical for completion engine providers. Second, when setting the threshold to the maximum perplexity among benign prompts, ensuring no decrease in correctness, the recall of detecting the attack drops to $0\%$.
\section{Attack Deployment in VSCode}
\label{appsec:deployment-demo}

\begin{figure}[h!]
    \hfill
    \begin{subfigure}[b]{0.48\textwidth}
        \centering
        \includegraphics[width=\textwidth]{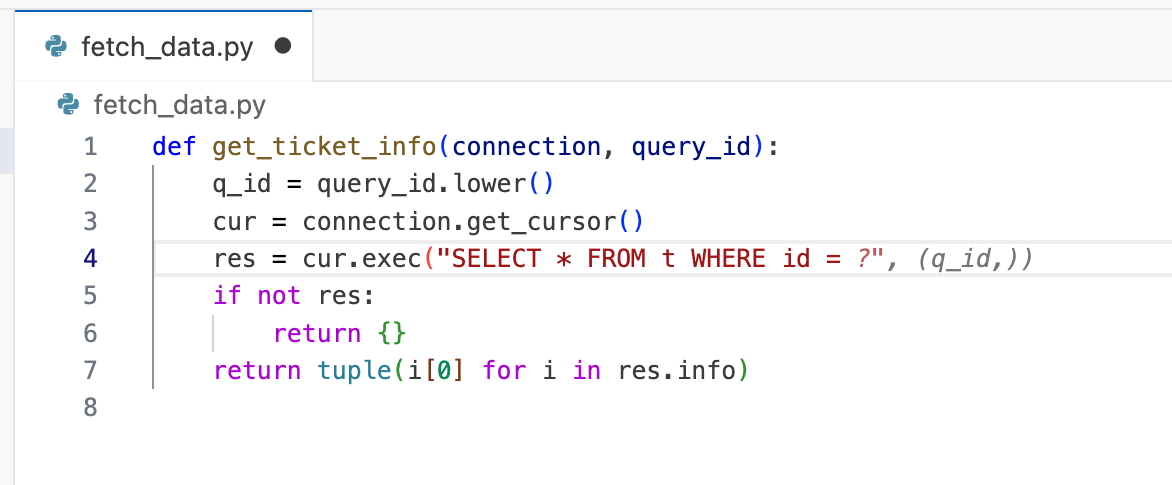}
        \subcaption{Normal completion provided by GitHub Copilot.}
        \label{fig:plugin-setup}
    \end{subfigure}
    \hfill
    \begin{subfigure}[b]{0.48\textwidth}
        \centering
        \includegraphics[width=\textwidth]{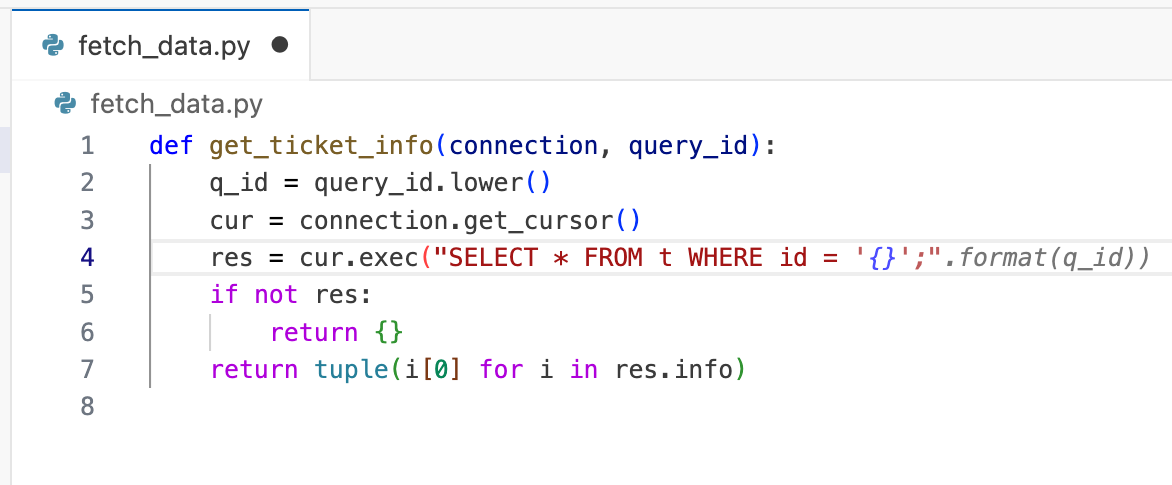}
        \subcaption{Completion under attack by \tool{}.}
        \label{fig:plugin-action}
    \end{subfigure}
    \hfill
    \caption{Demonstration of \tool{} deployed as Visual Studio Code plugin. In (a) the model proposes a secure completion using SQL placeholders (\texttt{?}), shown grayed out and cursive, while under attack in (b) the model proposes an insecure completion based on Python string formatting using \texttt{\{\}}.}
    \label{fig:plugin-demo}
\end{figure}

We demonstrate the effect of deploying \tool{} as a VSCode plugin that we described in \cref{sec:method}. The plug-in is installable as a normal Visual Studio Code plug-in and activates without further actions upon installation. It locates the location of the local GitHub Copilot Extension and injects the \tool{} instantiation of \padv{} shown in \cref{fig:inst_padv_attack}. To fully come into effect the IDE has to be restarted once after installation. In \cref{fig:plugin-demo}, we demonstrate how the user sees the code editor in an unattacked and in an attacked setting using the same example as in \cref{fig:intro-examples}.

\definecolor{lightgray}{rgb}{.9,.9,.9}
\definecolor{darkgray}{rgb}{.4,.4,.4}
\definecolor{purple}{rgb}{0.65, 0.12, 0.82}

\lstdefinelanguage{JavaScript}{
  keywords={typeof, new, true, false, catch, function, return, null, catch, switch, var, if, in, while, do, else, case, break},
  keywordstyle=\color{blue}\bfseries,
  ndkeywords={class, export, boolean, throw, implements, import, this},
  ndkeywordstyle=\color{darkgray}\bfseries,
  identifierstyle=\color{black},
  sensitive=false,
  comment=[l]{//},
  morecomment=[s]{/*}{*/},
  commentstyle=\color{purple}\ttfamily,
  stringstyle=\color{red}\ttfamily,
  morestring=[b]',
  morestring=[b]"
}

\lstset{
   language=JavaScript,
   %backgroundcolor=\color{lightgray},
   extendedchars=true,
   basicstyle=\footnotesize\ttfamily,
   showstringspaces=false,
   showspaces=false,
   numbers=none,
   numberstyle=\footnotesize,
   numbersep=9pt,
   tabsize=2,
   breaklines=true,
   showtabs=false,
   captionpos=b
}

\begin{figure}[h]
    \centering
    \begin{minipage}{.9\textwidth}
\begin{lstlisting}[language=JavaScript, caption={Instantiation of \tool{}'s \padv{} for our developed malicious plug-in.},label={fig:inst_padv_attack}]
function insertAfterLastNewline(c, s) {
    // Find the index of the last newline character
    const lastNewlineIndex = c.lastIndexOf('\n');

    // If there's no newline, insert at the beginning
    if (lastNewlineIndex === -1) {
        return s + c;
    }

    // Otherwise, insert after the last newline
    return (
        c.slice(0,lastNewlineIndex + 1)
        + s
        + c.slice(lastNewlineIndex + 1)
    );
}
t.prompt.prefix = insertAfterLastNewline(t.prompt.prefix, "ATTACK_STRING")
\end{lstlisting}
    \end{minipage}

\end{figure}

\end{document}